\documentclass[amsmath,amssymb,11pt]{article}
\usepackage{jheppub2}
\pdfoutput=1

\usepackage{amsmath,amssymb,amsthm,mathrsfs,bbm,bm}
\usepackage{latexsym,amscd,amsbsy,amsfonts,dsfont}
\usepackage{multirow}
\usepackage{graphicx}
\usepackage{xcolor}

\usepackage{caption}

\usepackage{empheq}
\newcommand*\widefbox[1]{\fbox{\hspace{.5em} #1 \hspace{.5em}}}

\newcommand{\beq}{\begin{equation}}
\newcommand{\eeq}{\end{equation}}
\newcommand{\beqa}{\begin{eqnarray}}
\newcommand{\eeqa}{\end{eqnarray}}
\newcommand{\bea}{\begin{eqnarray}}
\newcommand{\eea}{\end{eqnarray}}

\newcommand{\nn}{\nonumber}

\newcommand*{\affmark}[1][*]{\textsuperscript{#1}} % For the affiliations

\numberwithin{equation}{section}  % make eq labels (sec.num)
\allowdisplaybreaks

\title{General Effective Theories of Black Holes in the Large~$D$ Limit}

\author{Roberto Emparan\affmark[1,2],}
\emailAdd{emparan@ub.edu}

\author{Jordi Rafecas-Ventosa\affmark[2], and}
\emailAdd{jrafecasventosa@icc.ub.edu}

\author{Benson Way\affmark[3]}
\emailAdd{bway@umd.edu}

\affiliation{
\affmark[1]Institució Catalana de Recerca i Estudis Avançats (ICREA),
 Passeig Lluis Companys, 23, 08010 Barcelona, Spain\\
\affmark[2]Departament de Física Quàntica i Astrofísica  and Institut de Ciències del Cosmos, Universitat de Barcelona, Martí i Franquès, 1, 08028 Barcelona, Spain\\
 \affmark[3]Maryland Center for Fundamental Physics, University of Maryland, College Park, MD 20742, USA\\
}

\abstract{
We derive the general form of the effective equations governing black hole dynamics in the limit of a large number of dimensions $D$. These split into a universal \emph{soap-bubble} embedding condition for stationary configurations and a set of nonlinear dynamical evolution equations describing near-horizon fluctuations of $O(1/D)$ amplitude over horizon scales of $O(1/\sqrt{D})$. We obtain these equations in full generality, including arbitrary asymptotic sources in the near-horizon region, and we show that they form a parabolic system with a well-posed initial value problem. To connect the various approaches to large-$D$ black hole dynamics, we also show that both the embedding and dynamical equations can be derived from the covariant membrane formalism. We clarify the intrinsic scope of the large-$D$ approach, emphasizing that it yields a well-posed dynamical evolution only on horizon scales of $O(1/\sqrt{D})$, which is the range where the most relevant horizon dynamics occur. Our results highlight the versatility of these effective theories for studying a wide class of black hole phenomena. 
}

\begin{document}

\maketitle

%%%%%%%%%%%%%%%%%%%%%%%%%%%%%%%%%%%%%

\section{Introduction}
\label{sec:intro}

Over the past decade it has become clear that black hole dynamics greatly simplifies when studied under an expansion in the inverse of the spacetime dimension, $1/D$  \cite{Emparan:2013moa}. The simplification stems from two key features: first, the dynamics localize within a narrow region of thickness $\sim 1/D$ around the horizon; and second, the coupling between horizon fluctuations and gravitational radiation is suppressed by at least $\sim e^{-D}$, and is therefore non-perturbatively small. This decoupling isolates the horizon dynamics and makes it amenable to non-linear effective theories that are far simpler to solve than the full Einstein equations (see \cite{Emparan:2020inr} for a review).

Two complementary lines of development have emerged, which we denote as approaches H (for hydrodynamic) \cite{Emparan:2015hwa,Emparan:2015gva,Emparan:2016sjk} and M (for membrane) \cite{Bhattacharyya:2015dva,Bhattacharyya:2015fdk}. The distinction in nomenclature is not fundamental, as both incorporate hydrodynamic and elastic features, but reflects different structural choices, mainly in the balance between manifest covariance and gauge fixing.\footnote{The labels H and M are only suggestive: the H approach is not a gradient expansion and, unlike hydrodynamics, can naturally handle strictly finite wavelengths, while the M approach is only superficially related to the `old' membrane paradigm \cite{Damour:1978cg,Price:1986yy}, which lacked the structure of an effective theory where short-distance degrees of freedom are integrated out.} 

Approach H builds effective theories that, so far, have been tailored to specific classes of black holes. The approach relies on a concrete distinction between coordinates, but is practical and versatile: it applies to a wide range of problems, often yielding analytic or simple numerical solutions and frequently extending well beyond the original scope.\footnote{For a sample of the variety of applications, see for instance \cite{Herzog:2016hob,Rozali:2017bll,Herzog:2017qwp,Iizuka:2018zgt,Andrade:2018zeb,Andrade:2018nsz,Andrade:2019edf,Andrade:2019rpn,Suzuki:2021lrw,Emparan:2021ewh,Luna:2022tgh,Emparan:2023dxm,Jarvinen:2024vdi}.} Moreover, the approach can be readily adapted to different asymptotics or matter content.

Approach M aims for generality, formulating a fully covariant set of equations that governs the dynamics of black holes in Minkowski or (A)dS backgrounds. It has proved highly effective for extracting broad theoretical results, and reproduces many findings---such as quasinormal spectra and nonlinear effective equations for black branes---previously obtained by other methods \cite{Dandekar:2016fvw,Dandekar:2016jrp,Bhattacharyya:2016nhn,Bhattacharyya:2017hpj,Dandekar:2017aiv,Mandlik:2018wnw,Saha:2018elg,Dandekar:2019hyc,Biswas:2019xip,Saha:2020zho}.

\begin{figure}
\begin{center}
    \includegraphics[width=.7\textwidth]{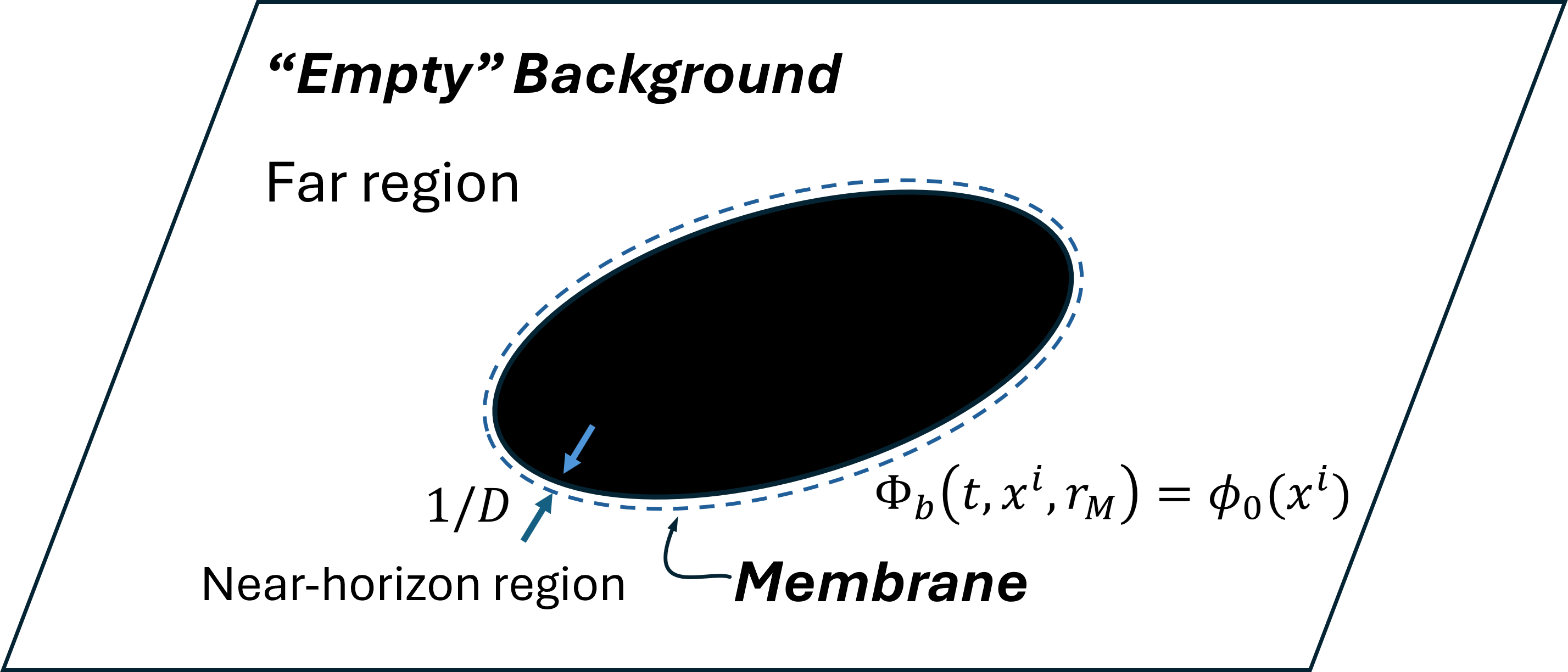}
\end{center}
\bigskip
    \includegraphics[width=.9\textwidth]{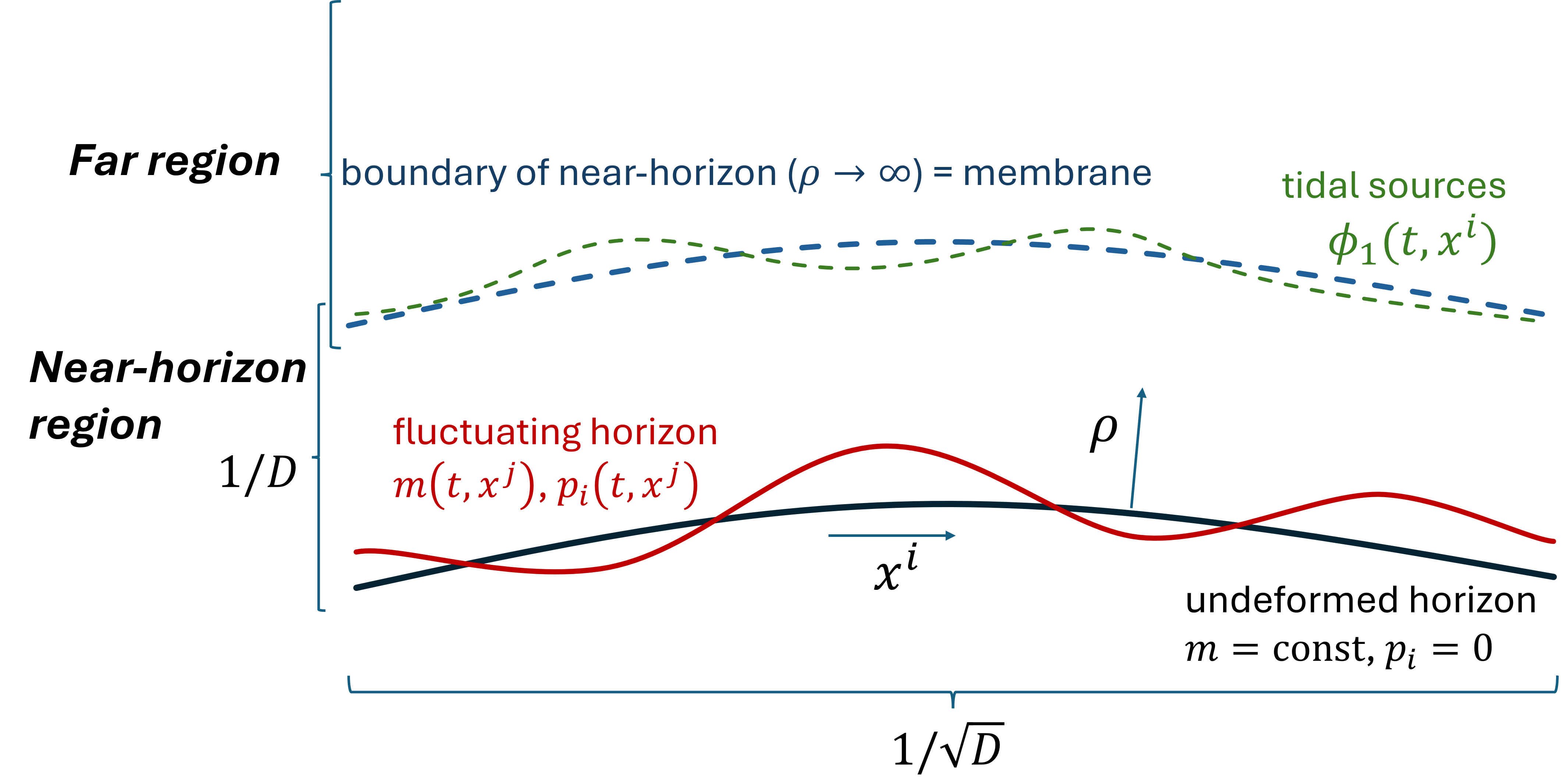}
\caption{\small Illustration of the main elements of the H approach (notation in Section \ref{sec:framework}). Top: The black hole is represented by a membrane— a surface in a background spacetime (Minkowski, (A)dS, or otherwise) defined by stationary embedding functions $\phi_0(x^i)$ satisfying the soap-bubble equation \eqref{eq:soapbubble}. The near-horizon region extends a distance $\sim 1/D$ from the horizon and is effectively thin for far observers. Bottom: Within this region and over horizon scales $O(1/\sqrt{D})$, the boundary admits deformations $\phi_1(t,x^i)$, while horizons fluctuations decaying as $e^{-\rho}$ are described by functions $m(t, x^j)$, $p_i(t, x^j)$. These obey the nonlinear, well-posed parabolic dynamical equations \eqref{eq:dyneqs}, whose explicit form \eqref{eq:effmp} is one of our main results. An undeformed horizon ($m=\textrm{const}$, $p_i=0$), requires $\phi_1(t,x^i)$ to satisfy a stationary soap-bubble condition at $O(1/D)$. The near and far regions overlap for matched asymptotics \cite{Asnin:2007rw,Emparan:2013moa}. The near-horizon region may also be treated as an autonomous system with freely chosen tidal sources $\phi_1(t,x^i)$ and without reference to any far-region embedding. Residual radial gauge transformations relate $\phi_1(t,x^i)$ to $m(t, x^j)$, $p_i(t, x^j)$.
} 
  \label{fig:membrane}
\end{figure}

Despite these advances, important questions in the large-$D$ program remain open. Chief among them are the detailed relation between the two types of effective theories and the precise range of black hole dynamics they can reliably capture. In this work, we aim to shed light on these issues, emphasizing the mutual consistency of the two approaches while at the same time extending their applicability to wider ranges of problems. Specifically, we will:
\begin{enumerate}
\item Explain that approach H naturally separates black hole dynamics into two components: (i) a ``soap-bubble'' equation that determines the shape of a \emph{stationary} horizon in a given background spacetime; and (ii) a set of dynamical equations %(with hydrodynamic-type variables) 
that govern time-dependent near-horizon fluctuations with amplitudes of order $1/D$.% and well-posed evolution 
\item Derive the general set of H-type effective large-$D$ equations for near-horizon dynamics, with well-posed evolution over horizon lengths of $O(1/\sqrt{D})$, in a Bondi-Sachs gauge but otherwise fully general.
\item Argue that effective equations for dynamics over horizon scales of $O(D^0)$, although sufficient to extract quasinormal spectra, do not furnish well-posed initial value problems.
\item Show how the M-type covariant equations reproduce both the soap-bubble equations for a black hole in Minkowski or (A)dS, and the H-type dynamical equations when applied to the Bondi-Sachs metrics introduced above.
\end{enumerate}
The key ideas of the H approach, along with the main elements of the large-$D$ effective theory explained in Section~\ref{sec:framework}, are illustrated schematically in Figure~\ref{fig:membrane}.

The main distinction between the H and M approaches is how the membrane directions enter the effective equations. In both cases there are directions that do or do not fluctuate along the horizon—‘active’ or ‘passive’. The former are a finite set, while the latter grow in number as $D\to\infty$, their only role being to localize the gravitational field near the horizon. 

The M-type equations unify the soap-bubble embedding and the near-horizon dynamics within a single, fully covariant framework that makes no explicit distinction between passive or active membrane directions. This covariance makes the formalism powerful for analyzing general properties of the membrane as an object in the background spacetime.

By contrast, in the H approach only the soap-bubble equation retains full covariance. The dynamical equations explicitly separate active and passive directions, and this makes them easy to apply in concrete settings. Moreover, these equations can be formulated allowing for arbitrary near-horizon asymptotics, which makes them effective for studying driven horizons, such as tidal deformations or holographic forced turbulence \cite{Andrade:2019rpn}.

Two important intrinsic limitations should be kept in mind regarding the types of time-dependent evolutions that the large-$D$ effective theories can describe:
\begin{itemize}
\item Time-dependent horizon fluctuations with amplitudes of $O(D^0)$ lie outside the regime of applicability, which only captures $O(1/D)$ variations in horizon size.
\item While the soap-bubble equation captures stationary horizon geometries on scales of $O(D^0)$ along the horizon, the near-horizon dynamical equations yield well-posed evolution only over horizon lengths of $O(1/\sqrt{D})$. Equations on $O(D^0)$ scales along the horizon can be obtained, but they fail to provide well-posed evolution.%\footnote{The examples we discuss here may likely be derived from the M equations.} 
\end{itemize}
The H approach makes these limitations transparent. We expect the M equations to be constrained in the same two ways, even if this is not manifest in their covariant formulation---a point of considerable interest for the broader large-$D$ program.

While our derivation of the H equations from the M formalism does not provide a rigorous proof of full equivalence, any residual differences appear marginal within the two limitations discussed above. Together with the soap-bubble embedding, we expect that the H equations should capture essentially all phenomena accessible to the covariant M formalism, insofar as the dynamical evolution problem is well-posed.\footnote{We anticipate that the equations for fluctuations over horizon lengths of $O(D^0)$, with ill-posed evolution, can also be derived from the M formalism.}

The article follows the structure of the points discussed above. Section~\ref{sec:framework} introduces the conceptual foundations of the large-$D$ limit and its intrinsic limitations. Section~\ref{sec:soapbub} reviews the soap-bubble equation. Section~\ref{sec:Hefftheory} derives the general H-type effective equations and discusses their main properties. Section~\ref{sec:examples} illustrates their use through examples. Section~\ref{sec:O1eqs} analyzes equations on $O(D^0)$ horizon scales, explains why they yield ill-posed evolution, and clarifies the contexts in which they remain useful. Section~\ref{sec:HfromM} rederives the H equations from the covariant M formalism. Section~\ref{sec:conclusion} summarizes our findings.

We occasionally review some earlier results to keep the presentation self-contained and to collect important elements of the large-$D$ approach scattered across the literature. In this sense, the article complements the broader review \cite{Emparan:2020inr}.

\section{Horizon shape and dynamics in the large-$D$ limit}\label{sec:framework}

\paragraph{Preliminaries.} In the limit of large spacetime dimension $D$, the gravitational field outside the horizon of a black hole decays exponentially in the radial direction. This behavior becomes manifest by examining the large-$D$ limit of the Schwarzschild-Tangherlini solution in Eddington–Finkelstein coordinates,
\begin{align}\label{eq:schlarge}
    ds^2 &=-\left(1-\frac{m}{r^{D-3}}\right) dt^2 +2 dt\, dr + r^2 d\Omega_{D-2}\nn\\
    &\to -\left(1-m e^{-\rho}\right) dt^2 + \frac{2}{D} dt\, d\rho +\left(1+\frac{2\rho}{D}\right) d\Omega_{D-2}+\dots\,,
\end{align}
where the radial coordinate $\rho$, related to the Schwarzschild area-radius by 
\begin{equation}\label{eq:nhrho}
    \rho= D\ln r\,, 
\end{equation}
is held fixed as $D\to\infty$. This coordinate naturally defines the near-horizon region, which has a thickness of order $1/D$ and extends from the horizon at $\rho = \ln m$ out to $\rho \to \infty$. 

As is customary in this context, we define
\begin{equation}\label{eq:nDp}
    n = D - p - 3\,, 
\end{equation}
where for finite $D$, $n$ appears as the exponent governing the radial fall-off of the gravitational field. We will interchangeably employ $1/D$ or $1/n$ as the small expansion parameter, since $p$ is kept fixed as $D \to \infty$. It labels the directions along which the horizon is allowed to fluctuate, which may be regarded as the spatial worldvolume directions of a black $p$-brane. 

To incorporate these directions $x^i$, $i=1,\dots,p$, we consider the metric of a boosted black $p$-brane,\footnote{The factors of $1/n$ that appear in the metric are dictated so that the terms that we show enter in the field equations at the same order in the expansion.}
\begin{align}\label{eq:bbrlarge}
    ds^2 = &-\left(1- e^{-\rho} m\right) dt^2 + \frac{2}{n} \left(dt- \frac{1}{n}\frac{p_i}{m} dx^i\right)d\rho -\frac{2}{n}e^{-\rho} p_i\, dt\, dx^i \nn\\
    &+\frac1{n} \left(\delta_{ij}+e^{-\rho}\frac{p_i p_j}{n m}\right) dx^i dx^j+\left(1+\frac{2\rho}{n}\right) d\Omega_{n+1}+\dots
\end{align}
This is obtained by boosting \eqref{eq:schlarge} with velocity $p_i/(n m)$ along approximately flat directions $x^i$, rescaled by $\sim 1/\sqrt{n}$ (more on this later). Since these velocities are $O(1/n)$, in the $n\to\infty$ limit  the boosts are non-relativistic and the system has Galilean invariance \cite{Emparan:2016sjk}.

Starting from this point, Ref.~\cite{Emparan:2015gva} derived the H-type effective dynamical equations by promoting $m$ and $p_i$ to functions of $(t, x^i)$ and determining the conditions that these “horizon collective coordinates” must satisfy for the Einstein equations to hold order by order in the $1/n$ expansion. The resulting equations are
\begin{subequations}\label{eq:mpeq}
    \begin{align}
    \partial_t m -\partial_i\partial^i m &=-\partial_i p^i\,, \label{eq:meq}\\
    \partial_t p_i -\partial_j\partial^j p_i &= \partial_i m -\partial^j\left(\frac{p_i p_j}{m} \right)\,.\label{eq:peq}
\end{align}
\end{subequations}

As shown in \cite{Emparan:2016sjk}, if we change $p_i=m v_i +\partial_i m$ then these take the form of hydrodynamic continuity equations for the mass and momentum of a fluid with velocity field $v_i$. However, we will mostly use the variable $p_i$ since it slightly simplifies the equations and much of their analysis.

We emphasize that the metric \eqref{eq:bbrlarge} describes only the near-horizon region of a black hole or black brane, extending over a radial distance of order $\sim 1/D$ away from the horizon, and over lengths $\sim 1/\sqrt{D}$ along the horizon. Its asymptotic limit, or boundary, lies at $\rho \to \infty$ (but $\rho/D\ll 1)$; beyond this begins the far region, where the geometry approaches flat Minkowski spacetime exponentially fast in $D$. The generalization to an AdS black brane simply results in a reversal of the sign of $\partial_i m$ in \eqref{eq:peq}---we will rederive these equations in Section~\ref{subsec:eqsbbranes}.

\paragraph{General H approach.}
An important feature of the H approach is that it naturally extends the effective theory to a wide class of background spacetimes in which the black hole is embedded. To explain this, we introduce a schematic notation that suppresses indices and collectively denotes metric components by $\Phi(t, x^i, r)$, with $r$ the radial coordinate orthogonal to the horizon.
The near-horizon expansion is then expressed as
\begin{equation}\label{eq:genans}
    \Phi(t,x^i,r) = \phi(t,x^i) %+\frac{1}{D}\varphi(t,x^i)
    +e^{-\rho} \psi(t,x^i)\,.
\end{equation}
Here $\psi(t, x^i)$ denotes the functions $m(t, x^i)$, $p_i(t, x^k)$, and combinations such as $p_i p_j/m$ that appear in $g_{tt}$, $g_{ti}$, and $g_{ij}$ in \eqref{eq:bbrlarge}. These $\psi(t, x^i)$ may include factors of $1/D$, depending on the metric component---we will be much more precise later. The key difference with \eqref{eq:bbrlarge} is that the data at the boundary $\rho\to\infty$ of the near-horizon region, 
\begin{equation}
    \phi(t,x^i) = \phi_0(t,x^i)+\frac1{D}\phi_1(t,x^i)+\dots
\end{equation}
% and $\varphi(t, x^i)$
are no longer constant, but allowed to vary along the horizon directions.

This makes it possible to describe more general ways in which the near-horizon region is glued to the far-zone background. Let the background—Minkowski, AdS, or otherwise—be given by metric functions $\Phi_b(x^\mu)$. Choose a surface $M$ in this background and introduce Gaussian coordinates $(t, x^i, r)$ such that $M$ is at fixed $r$. To leading order in $1/D$ we impose
\begin{equation}
    \Phi_b(t, x^i, r_M) = \phi_0(t,x^i)\,,
\end{equation}
so that \emph{the boundary of the near-horizon region is represented by the surface $M$}. We refer to $M$ as the ``black hole membrane,'' or simply the membrane.\footnote{We avoid the term “horizon membrane” because it lies not at the actual horizon, but at the interface where the near-horizon region meets the background, a distance $\sim 1/D$ away.}

Not every surface in the background qualifies as a black hole membrane, since its shape is ultimately fixed by the Einstein equations. Requiring the ansatz \eqref{eq:genans} to solve them at leading order yields:
\begin{enumerate}
    \item $\partial_t \phi_0(t,x^i)=0$: the membrane is stationary.\footnote{We assume a static background with Killing vector $\partial_t$, though it would be interesting to investigate if this could be relaxed. }
    \item The embedding $\phi_0(x^i)$ satisfies the
    \begin{equation}\label{eq:soapbubble}
        \textsl{Soap-bubble equation}:\quad  \left.\gamma^{-1} K\right|_M=2\kappa_g\,,
    \end{equation}
    where $K$ is the trace of the extrinsic curvature of $M$, and $\gamma^{-1}=\sqrt{-g_{tt}(1-v^2)}$ measures the local gravitational and Lorentz redshifts along the membrane. %\footnote{Defined relative to asymptotically static observers.} 
    The constant $\kappa_g$ is the black hole surface gravity. 
\end{enumerate}
These results were derived in \cite{Emparan:2015hwa} and \cite{Suzuki:2015iha}, who found that condition (1) is generally sufficient to obtain a consistent set of equations when $D\to\infty$, though they did not prove its necessity in full generality. Detailed studies indicate the possibility of mildly time-dependent embeddings, but these appear to be of marginal physical relevance.

Solving the Einstein equations one higher order in $1/D$ implies:
\begin{enumerate}
\item The variables $m(t,x^i)$ and $p_i(t,x^j)$ obey the
\begin{align}\label{eq:dyneqs}
    \textsl{Dynamical effective equations}:\quad  \partial_t m&=\nabla^2 m+\mathcal{F}(m,p_j,\phi)\,,\nn \\ \partial_t p_i&=\nabla^2 p_i+\mathcal{F}_i(m,p_j,\phi)\,,
    %\mathcal{D}[\psi(t,x^i)] = F\left(\varphi(t,x^i)\right)\psi(t,x^i)\,,
\end{align}
where $\mathcal{F}$ and $\mathcal{F}_i$ contain lower-order derivatives of $m$, $p_i$, in addition to $\phi$. The function $\mathcal{F}$ is linear in $m$ and $p_i$, while $\mathcal{F}_i$ contains a single nonlinear term of the form in \eqref{eq:peq}.

\item The near-horizon boundary admits \emph{arbitrary} tidal sources $\phi_1(t,x^i)$ that deform and drive the horizon. If we require that the horizon is uniform and stationary, these sources are constrained to solve the soap-bubble equation \eqref{eq:soapbubble} at $O(1/D)$.
\end{enumerate}

The equations \eqref{eq:dyneqs}, which extend \eqref{eq:mpeq} by incorporating generic sources, describe general fluctuations of the horizon with amplitudes $1/D$. Two crucial points of these equations are:
 \begin{itemize}
     \item The spatial coordinates $x^i$ have been rescaled by a factor $1/\sqrt{D}$ relative to the far region, so that the effective dynamics is confined to horizon lengths of $O(1/\sqrt{D})$.
     \item The second-order spatial derivatives of $m$ and $p_i$ ensure that the system defines a well-posed parabolic evolution problem.
 \end{itemize} 
The two issues are closely related: equations similar to \eqref{eq:dyneqs} that are valid on horizon scales of $O(D^0)$ can be written, but they involve only first-order spatial derivatives of $m$ and $p_i$, rendering the initial value problem ill-posed. They can still be used to find stationary configurations, effectively reducing to $O(1/D)$ soap-bubble equations, and (with some caveats) compute quasinormal spectra. 

We emphasize that the way we decompose \eqref{eq:genans} is not fully gauge-invariant: mixing the radial and $x^i$ coordinates can remove or introduce some of the sources $\phi_1$, while modifying $m$ and $p_i$ (see Section~\ref{subsec:membend}). Although this gauge dependence could in principle be fixed, it is in fact a useful feature rather than a drawback: the resulting freedom allows the same black hole to be described by different---but physically equivalent---effective theories, each adapted to a different calculational or conceptual purpose, as illustrated in Section~\ref{sec:examples}.

Finally, the method highlights a close conceptual parallel with AdS/CFT: near-horizon dynamics decouple from the ambient spacetime \cite{Emparan:2014aba}. In this picture, the boundary geometry of the near-horizon region acts as a gravitational source that perturbs the black hole, independent of any far-region embedding. This viewpoint has already been used to model holographic forced turbulence \cite{Andrade:2019rpn}, and it can likewise be employed to compute tidal Love numbers in the large-$D$ limit.

In the next three sections we derive the explicit form of the equations \eqref{eq:soapbubble} and \eqref{eq:dyneqs}, discuss their key properties, and illustrate them with representative examples.

\section{Soap-bubble equation and solutions}\label{sec:soapbub}

We begin with a brief review of the soap-bubble condition derived in \cite{Emparan:2015hwa,Suzuki:2015iha}.

The equation
\begin{equation}\label{eq:staticsb}
    \left.\sqrt{-g_{tt}} K\right|_M=2\kappa_g
\end{equation}
 describes \emph{static} embeddings of the black hole as a membrane at a surface $M$ in a background, in the leading large-$D$ limit of Einstein gravity with possibly a cosmological constant. Upon extending the solution into the near-horizon region, the constant $\kappa_g$ is identified as the surface gravity of the Killing generator of the horizon, $\partial_t$.

This equation generalizes to stationary black holes, still in a static background, in the form \eqref{eq:soapbubble}.
The background admits a timelike Killing vector $\partial_t$, but more generally, it may also possess another Killing vector $k$ that is timelike on the membrane, along which one can define a unit-normalized velocity 
\begin{align}
    u =\frac{k}{|k|}\,,
\end{align}
which we take to be tangent to the membrane. Then, $\gamma^{-1}=|k|$ combines the gravitational and Lorentz redshifts on the membrane. To see this in the relevant case of angular rotations, where
\begin{align}\label{eq:rotkilling}
    k=\partial_t+\Omega \partial_\phi
\end{align}
with constant $\Omega$, we define the radius
\begin{equation}
    R(x^i)=\left.\sqrt{\frac{g_{\phi\phi}}{-g_{tt}}}\right|_M\,.
\end{equation}
Then \eqref{eq:soapbubble} becomes
\begin{equation}\label{eq:station2sb}
     \left.\sqrt{-g_{tt}(1-\Omega^2 R^2)}\,K\right|_M=2\kappa_g\,.
\end{equation}
This equation determines the surface gravity $\kappa_g$ and angular velocity $\Omega$ of the black hole horizon in the large $D$ limit.

For illustration, we discuss two solutions studied in \cite{Emparan:2015hwa,Suzuki:2015iha}.

\paragraph{Schwarzschild and Myers-Perry black holes as soap bubbles.} A straightforward calculation in the background
\begin{equation}
    ds^2=-\left(1+\sigma\frac{r^2}{L^2}\right)dt^2 + \frac{dr^2}{1+\sigma\frac{r^2}{L^2}}+ r^2 d\Omega_{D-2}\,,
\end{equation}
 (which is (A)dS or Minkowski for $\sigma=\pm 1, 0$)
shows that the spherical surface $r=r_0$ satisfies \eqref{eq:staticsb} with
\begin{equation}
    \kappa_g= \frac{D}{2r_0}\left(1+\sigma\frac{r_0^2}{L^2}\right)+O(D^0)\,.
\end{equation}
This matches the surface gravity of the Schwarzschild-(A)dS black hole at large $D$, which is therefore described as a spherical soap bubble.

Stationary rotating black holes can be obtained in a similar way. For instance, writing Minkowski space in spheroidal coordinates,
\begin{equation}
    ds^2=-dt^2 +(r^2+a^2\cos^2\theta)\left(\frac{dr^2}{r^2+a^2}+d\theta^2\right)+(r^2+a^2)\sin^2\theta d\phi^2 +r^2\cos^2\theta\, d\Omega_{D-4}\,,
\end{equation}
the extrinsic curvature of the surface $r=r_0$ is
\begin{equation}
    K=\frac{D}{r_0}\sqrt{\frac{r_0^2+a^2}{r_0^2+a^2\cos^2\theta}}+O(D^0)\,.
\end{equation}
Consider now a Killing vector of the form \eqref{eq:rotkilling}, so that $ \gamma^{-1}=\sqrt{1-\Omega^2 R^2}$ with
\begin{align}
    R^2=\left.g_{\phi\phi}\right|_{r_0}=(r_0^2+a^2)\sin^2\theta\,.
\end{align}
To satisfy \eqref{eq:station2sb} we must set
\begin{align}
     \Omega=\frac{a}{r_0^2+a^2}\,,\qquad \kappa_g=\frac{D}{2 r_0}\,.
\end{align}
These reproduce the correct surface gravity and angular velocity of the Myers–Perry black hole at large $D$ \cite{Myers:1986un,Emparan:2013moa}. The rotating black hole is thus recovered as a spheroidal soap bubble.

In Section~\ref{sec:examples} we will see how these embeddings can induce sources in the near-horizon geometry.

\section{General H-type effective equations }\label{sec:Hefftheory}

We now proceed to derive a general set of effective equations for the near-horizon dynamics by solving the Einstein equations
% \begin{equation}
% R_{MN} - \frac{2\Lambda}{D-2} g_{MN}, 
% \qquad 
% \Lambda = \sigma \frac{(D-1)(D-2)}{2L^2},
% \end{equation}
\begin{equation}\label{eq:genEins}
R_{MN} = \sigma\frac{D-1}{L^2} g_{MN}
\end{equation}
to next to leading order in $1/D$. We consider positive, negative or zero cosmological constant,
\begin{equation}
    \sigma=\pm 1, 0\,. 
\end{equation}
When $\sigma=\pm 1$ we will fix the length units by setting $L=1$. Typically the solutions will contain another length $r_0$ that sets the horizon size. When $\sigma=0$ we can choose it to be $r_0=1$, but when $\sigma=\pm 1$, it remains as a parameter which should be understood as $r_0/L$.

\subsection{Setup}

We take an ingoing Bondi-Sachs ansatz for the metric
\begin{equation}\label{eq:BondiSachs}
ds^2 = G_{ij}(X)\,dx^i dx^j + W(X)\, d\Sigma_{n+1}^{(\kappa)}
+ 2 dt \Bigl( dr - A(X) dt - F_i(X) dx^i \Bigr),
\end{equation}
where $X = (t, \vec{x}, r)$,  $\vec{x}$ has dimension $p$ (see \eqref{eq:nDp}), and $\Sigma_{n+1}^{(\kappa)}$ is an $n+1$-dimensional space of constant curvature equal to \begin{equation}
    \kappa=\pm 1, 0\,.
\end{equation} 

Other than assuming the presence of the cyclic factor $\Sigma_{n+1}^{(\kappa)}$ in the geometry, this ansatz is fully general. In Appendix~\ref{app:BondiSachseqns} we derive the form of the Einstein equations for these metrics with arbitrary $n$.

A residual gauge freedom remains in the choice of radial coordinate, with transformations of the form
\begin{equation}\label{eq:rgauge}
    r\to r + H(t,\vec{x})
\end{equation}
that preserve the ansatz. This freedom will be a useful feature of the formalism.

\subsection{Scalings with $D$ and well-posed evolution}\label{subsec:scaling}

The expansion in $1/D$ acquires meaning once we specify how the various quantities in \eqref{eq:BondiSachs} scale with $D$. To this end, we first zoom in on the near-horizon region by requiring that the radial coordinate $\rho$ introduced in \eqref{eq:nhrho} remain fixed as $D\to\infty$. This implies that, locally, the amplitude of the fluctuations that we describe is $O(1/D)$.

To determine the scaling of the metric functions in \eqref{eq:BondiSachs}, we examine their behavior in the black brane solution \eqref{eq:bbrlarge}. We see that $A$ and $W$ are of $O(D^0)$. In contrast, $G_{ij}$ is of $O(1/D)$, since we rescaled the spatial worldvolume directions and thus focused on a region along the horizon of length $O(1/\sqrt{D})$. In a black brane, this is justified because this is the characteristic wavelength of sound modes (for an AdS black brane) or Gregory-Laflamme unstable modes (for an AF black brane). 

Nevertheless, could there be effective dynamical equations that capture fluctuations on horizon scales $O(D^0)$? 
Such equations can indeed be written down, and Section~\ref{sec:O1eqs} presents some examples. However, they do not define well-posed evolution because, as we will presently argue, they are schematically of the form
\begin{equation}\label{eq:parabol}
    \partial_t^2 \psi(t,x)+ A(x)\, \partial_x \psi(t,x) = \textrm{(lower order terms)}\,.
\end{equation}
This is a parabolic system with space and time effectively exchanged, and the absence of second-order spatial derivatives causes high-frequency modes to grow uncontrollably. In momentum space, with $\partial_x\to ik$, if we take $|k|\gg 1$ we can approximate $A$ as constant and neglect the lower-order terms, giving modes
\begin{equation}\label{eq:parabol2}
\psi \sim \exp\left((1+i)\sqrt{\frac{A |k|}{2}}t\right)\,,
\end{equation}
which reach $O(1)$ amplitude in arbitrarily short times for sufficiently large $|k|$. This violates continuous dependence on initial data, so the evolution is ill posed.\footnote{Equations of the type \eqref{eq:parabol} may still admit physically meaningful solutions, such as boson stars or oscillatons \cite{Rozali:2018yrv}.}

\paragraph{Degeneration in the large $n$ limit.} Although we do not have a general proof that \eqref{eq:parabol} is a feature of all large-$D$ effective equations with $x$-variations of $O(D^0)$, we can provide a simple, broadly applicable argument for the suppression of second-order spatial derivatives.%\footnote{This extends considerations in appendix C of \cite{Suzuki:2015iha} and appendix B of \cite{Andrade:2018nsz}.}

To unclutter the notation, we consider only one direction $x^i = x$ along the horizon. Then the metric \eqref{eq:BondiSachs} contains a factor
\begin{equation}\label{eq:ds2H}
    dx^2 +W(x) \,d\Sigma_{n+1}^{(\kappa)}\,.
\end{equation}
We take, as a proxy for the metric functions, a massless scalar field $\chi$ in a spacetime containing this factor. Separating the $x$ variable in the wave equation will yield
\begin{equation}\label{eq:chieq}
    \nabla^2\,\chi (x) \equiv \partial_x^2\,\chi +(n+1)\frac{W'}{2W} \partial_x\chi = \Lambda \chi (x)\,,
\end{equation}
where $\nabla^2$ is the Laplacian in \eqref{eq:ds2H} and $\Lambda$ is the separation constant. For instance, if \eqref{eq:ds2H} described a sphere $S^{n+2}$, with $x$ a polar angle and $W(x)=\cos^2 x$, this would be the equation for spherical harmonics in $S^{n+2}$, with $\Lambda =\ell(n+\ell+1)$.
Taking in general $\Lambda = n\lambda + O(n^0)$,
then
\begin{equation}\label{eq:red1}
    \frac{W'}{2W}\,\partial_x\chi = \lambda \chi+O(1/n)\,.
\end{equation}
Thus, the second-order spatial derivatives become suppressed when $n\to\infty$ and the system reduces to first-order in $x$. This degeneration of the principal symbol of \eqref{eq:chieq} is the reason for the character of \eqref{eq:parabol}. %\footnote{For AF black branes}
As discussed in \cite{Andrade:2018nsz}, it has a physical counterpart in the loss of information about the structure of the solutions in the $x$ direction---in particular, the absence of fluctuations that propagate along the horizon.\footnote{When the finite-$n$ problem is elliptic, the degeneration of the principal symbol in the limit $n\to\infty$ need not lead to pathologies, but can actually result in useful simplifications \cite{Emparan:2019obu}.}

The second-order spatial derivatives and a finer structure of horizon dynamics can be retained if we focus on a narrow region of extent $\sim 1/\sqrt{n}$ around an extremum of $W$. Taking this to be at $x=0$, so that $W'(0)=0$ and
\begin{equation}
    W(x)=W_0\left(1 + c\, x^2 + O(x^3)\right)\,,
\end{equation}
with $c=W''(0)/(2 W(0))$,
we rescale
\begin{equation}
    x\to \frac{x}{\sqrt{n}}\,,
\end{equation}
so \eqref{eq:ds2H} becomes
\begin{equation}\label{eq:nearmax}
    \frac{dx^2}{n} +W_0\left(1+c\,\frac{x^2}{n}\right) d\Sigma_{n+1}^{(\kappa)}\,.
\end{equation}
Now \eqref{eq:chieq} yields
\begin{equation}\label{eq:red2}
    \partial_x^2\chi +c\, x\partial_x\chi = \lambda\chi+O(1/n)\,,
\end{equation}
which is of second order in $x$. When either first or second derivatives in time are included together with \eqref{eq:red2} in the full equation, we can obtain a good evolution equation with suppressed high-frequency modes.

We will see in Section~\ref{sec:O1eqs} that equations with $O(D^0)$ range over the horizon can still be useful to study stationary configurations and their quasinormal modes---the kind of problems that have been mostly tackled with the M formalism---but not for well-defined time evolution.

\subsection{Effective Equations}

In fixing the structure and $D$-scaling of the metric functions, we heed the arguments above and restrict the near-horizon region to lengths of order $1/\sqrt{D}$. 

To isolate the fluctuations in the overall size $W$ of the space $\Sigma_{n+1}^{(\kappa)}$ we write it as
\begin{equation}
W = r_0^2 r^2 S(t,\vec{x},\rho)^2\,,
\end{equation}
and for the other metric functions we set 
\begin{subequations}\label{eq:sources}
\begin{align}
S(t,\vec{x},\rho) &= 1 + \frac{s(t,\vec{x})}{D}
+ \frac{1}{D^2}\frac{S^{(1)}(t,\vec{x},\rho)}{\lambda^2} 
+ O\!\left(\frac{1}{D^3}\right), \\
A(t,\vec{x},\rho)&=\frac{1}{2}\lambda^2\left(1-e^{-\rho}m(t,\vec{x})\right)+\frac{1}{D}A^{(1)}(t,\vec{x},\rho)+O\!\left(\frac{1}{D^2}\right)\\
    F_i(t,\vec{x},\rho)&=\frac{1}{D}\left(f_i(t,\vec{x})+e^{-\rho}p_i(t,\vec{x})\right)+\frac{F^{(1)}_i(t,\vec{x},\rho)}{\lambda^2 D^2}+O\!\left(\frac{1}{D^3}\right)\\
G_{ij}(t,\vec{x},\rho) &\sim \frac{1}{D}\left[\gamma_{ij}(t,\vec{x}) 
+ \tfrac{1}{D}\frac{G^{(1)}_{ij}(t,\vec{x},\rho)}{\lambda^2%\frac{\kappa}{r_0^2} - \sigma
} 
+ O\!\left(\tfrac{1}{D^2}\right)\right]\,,
\end{align}
\end{subequations}
where $G_{ij}$ and $F_i$ have been chosen to be $O(1/D)$ for the reason explained in the previous subsection. We have conveniently defined
\begin{equation}\label{eq:deflambda}
    \lambda^2 =\frac{\kappa}{r_0^2}-\sigma\,,
\end{equation}
and we will require $\lambda^2>0$. This holds in all the most relevant cases: Minkowski ($\sigma=0,\kappa=1$), global and Poincaré-AdS  ($\sigma=-1,\,\kappa=1,0$), and the static patches of hyperbolic AdS and dS ($\sigma=-1,+1,\,\kappa=-1$). In particular, in Poincaré-AdS and in Minkowski backgrounds (with $r_0=1$) we simply have $\lambda^2=1$. Appendix \ref{app:rescSigma} explains how these and other useful backgrounds arise within this ansatz.

Note that to organize the $1/D$ expansion, it is not the metric itself---whose components naively mix different orders---that should be examined, but rather the way the metric functions enter the equations of motion. For instance $S(t,\vec{x},\rho)$ enters in the equations exponentiated to $n+1$, thus rendering $s$ a leading-order term like $f_i$ or $\gamma_{ij}$.

This form of the metric already satisfies the lowest-order terms of each component of the Einstein equations \eqref{eq:Einseqs}, with $s(t,\vec{x})$, $f_i(t,\vec{x})$, and $\gamma_{ij}(t,\vec{x})$ remaining arbitrary functions, which thus act as asymptotic sources. Note that there is no arbitrariness or $(t,\vec{x})$-dependence in the asymptotic form of the $g_{tt}$ component $A$; this follows from our gauge choices. A different radial gauge could yield varying sources in $A$. Even within this ansatz, the definition of the asymptotic sources retains a residual gauge dependence through the transformations \eqref{eq:rgauge}, whose implications we will analyze in the next subsection.

In the effective equations, all indices are raised and lowered with the spatial metric $\gamma_{ij}$, from which we also define the usual covariant derivative $\nabla_i$ and Ricci tensor $R_{ij}$.

To proceed, we substitute this ansatz into the Einstein equations~\eqref{eq:Einseqs} and expand them in powers of $1/D$.\footnote{The calculations were performed using \textsl{Mathematica}.}
At the next order in $1/D$, we can integrate the radial dependence in $G^{(1)}_{ij}$ and $S^{(1)}$ and impose regularity at the horizon. We obtain
\begin{subequations}\label{eq:nextsol}
\begin{align}
G^{(1)}_{ij}(t,\vec{x},\rho) &= e^{-\rho}\frac{p_i p_j}{m}
- 2\rho\Big[\nabla_i\nabla_j s - R_{ij} + \tfrac{1}{2}\partial_t \gamma_{ij} 
+ \sigma \gamma_{ij} + \nabla_{(i} f_{j)} \Big] + g^{(1)}_{ij}(t,\vec{x}), \\
S^{(1)}(t,\vec{x},\rho) &= -\frac{1}{2} e^{-\rho} \frac{p^2}{m} 
+ \rho\, s^{(1)}_a(t,\vec{x}) + s^{(1)}_b(t,\vec{x}), 
\end{align}
\end{subequations}
where $s^{(1)}_a$, $s^{(1)}_b$, and $g^{(1)}_{ij}$ are integration functions. Let us examine the terms that grow with $\rho$. Some of these---involving $s$, $f_i$, and $\gamma_{ij}$---are fixed by the asymptotic sources. In addition, there appears a new integration function $s^{(1)}_a$ that will act as an extra source in the effective equations. However, these effective equations will not depend on $s^{(1)}_b$ or $g^{(1)}_{ij}$, which do not grow towards the boundary $\rho\to\infty$.

Additional equations appear for the higher-order terms $A^{(1)}$ and $F_i^{(1)}$, but solving them is unnecessary for deriving the dynamics---their complete form is needed if one wishes to reconstruct the spacetime metric at this order, or to obtain $1/D$ corrections to the effective equations, which we will not do. Nevertheless, we give the asymptotic behaviors of their solutions in the region $\rho\to\infty$, as they will be required in Section.~\ref{subsec:soap1D}. These are
\begin{subequations}\label{eq:nextAF}
\begin{align}
    -A^{(1)}&\rightarrow \sigma\rho+\left(\lambda^2+\sigma\right)s+s_a^{(1)}+\frac{1}{2}f^2+\frac{1}{2}(\nabla s)^2+f\cdot\nabla s+\partial_t {s}+\frac{1}{2}R\,,\label{eq:A1asym}\\
    F^{(1)}_j &\rightarrow 2\rho\Biggl(\nabla^i\left(\nabla_{[j}f_{i]}-R_{ij}-\frac{\partial_t{\gamma}_{ij}}{2}\right)-\left(f^i+\nabla^i s\right)\left(\nabla_j\nabla_i s+\frac{\partial_t{\gamma}_{ij}}{2}+\nabla_{(j}f_{i)}\right)\nn\\* & \qquad\quad +\tfrac{1}{2}\nabla_j\text{tr}(\partial_t \gamma) \Biggr).
\end{align}
\end{subequations}

At this order in $1/D$ we also obtain a set of equations for $m$ and $p_i$ that are independent of $\rho$: these constitute the dynamical effective equations. 
To simplify them, we set, without loss of generality,
\begin{equation}
f_0(t,\vec{x}) = s^{(1)}_a(t,\vec{x})+\nabla^2 s + (\nabla s)^2 + 2(\lambda^2+\sigma) s
- \tfrac{1}{2}\text{tr}(\partial_t \gamma) + \tfrac{1}{2} f^2 
+ f\cdot\nabla s, 
\end{equation}
which essentially replaces $s^{(1)}_a$ by $f_0$. Furthermore, we define
\begin{equation}
\hat{p}_i \equiv p_i + m f_i. 
\end{equation}
The resulting effective equations can then be written as
\begin{subequations}\label{eq:effmp}
\begin{empheq}[box=\widefbox]{align} %\begin{align}
\Big(\partial_t + \tfrac{1}{2}\,\text{tr}(\partial_t \gamma) + \partial_t s \Big) m 
&= \Big( \nabla^i + \nabla^i s \Big)(\nabla_i m - \hat{p}_i), \label{eq:effm}\\[6pt]
\Big(\partial_t + \tfrac{1}{2}\,\text{tr}(\partial_t \gamma) + \partial_t s \Big) \hat{p}_j 
&= \Big( \nabla^i + \nabla^i s \Big)\!\left(\nabla_i \hat{p}_j - \frac{\hat{p}_i \hat{p}_j}{m} + m\,\partial_t \gamma_{ij}\right) \nonumber \\[6pt]
&\quad + \big( 2\nabla^i m - \hat{p}^i \big)\big(\nabla_i \nabla_j s - R_{ij}\big) \nonumber \\[6pt]
&\quad + \left(\lambda^2+2\sigma\right)%\Big(\frac{\kappa}{r_0^2} + \sigma \Big)
\nabla_j m 
- 2(\nabla^i m - \hat{p}^i)\nabla_{[i} f_{j]} \nonumber  \\[6pt]
&\quad
+ (\partial_t f_j + \nabla_j f_0)\,m. \label{eq:effp}
\end{empheq} %\end{align}
\end{subequations}
These equations determine the mass and momentum density profiles, $m$ and $\hat{p}_i$---or the local height and velocity of the horizon---for specified sources $s$, $f_0$, $f_i$, $\gamma_{ij}$. They are one of the main results in this article. 

\subsection{Sources, Symmetries, and Gauges} \label{subsec:membend}

\paragraph{Meaning of sources.} We can see that $f_0$ and $f_i$ play, respectively, the role of a gravitational potential yielding a Newtonian force $m\nabla_i f_0$, and of a background velocity field which can drag and accelerate the brane through $m\partial_t f_i$, or induce vorticity through $\nabla_{[i} f_{j]}$. In accordance with the equivalence principle, the acceleration field $\partial_t f_i$ and the gravitational force field $\nabla_j f_0$ enter multiplying the mass density in exactly the same way---a point to which we will return below.

The spatial metric $\gamma_{ij}$ gives rise to tidal effects through its time variation, $\partial_t \gamma_{ij}$, and its spatial curvature $R_{ij}$. 

The role of $f_0$ differs from that of $s$, $\gamma_{ij}$, and $f_i$. The latter enter as background values at infinity in the leading non-trivial order of the metric functions \eqref{eq:sources}. By contrast, $f_0$ (essentially $s_a^{(1)}$) appears as the coefficient of a correction of order $\rho/D$ in \eqref{eq:nextsol}. As we will see in Section~\ref{subsec:soap1D}, this makes $f_0$ proportional (up to a Lorentz redshift) to the surface gravity of the black hole at $O(1/D)$.

The function $s$ encodes the bending of the membrane along the directions of $\Sigma_{n+1}^{(\kappa)}$, over lengths $O(1/\sqrt{n})$. It is not, strictly speaking, an external source: as we will show next, the way this bending is represented depends on the choice of radial gauge.

\paragraph{Membrane-bending symmetry.} 
The Bondi-type metric ansatz admits residual radial gauge transformations of the form \eqref{eq:rgauge}. In near-horizon variables, this corresponds to shifting
\begin{equation}\label{eq:rshift}
    \rho\to \rho+h(t,\vec{x})
\end{equation}
with arbitrary $h(t,\vec{x})$. Under this transformation, the fields of the effective theory change as
\begin{subequations}\label{eq:fulltransf}
    \begin{align}
    & m\to e^hm\,, \qquad p_i \to e^hp_i\,,\qquad  f_i\to f_i+\nabla_i h\,,\label{eq:transmpfi}\\
    & f_0\to f_0 -(\lambda^2+2\sigma)h-\nabla^2 h +\frac12 (\nabla (s-h))^2-\frac12 (\nabla s)^2\,,\\
    &s\to s-h\,.
\end{align}
\end{subequations}
To see that the effective equations \eqref{eq:effmp} retain their form under these transformations, it is convenient to note that 
\begin{equation}\label{eq:phtrans}
    \hat{p}_i\to e^h\left(\hat{p}_i+m\nabla_i h\right)\,,    
\end{equation}
and to use the Ricci tensor identity
\begin{equation}
    R_{ij}\nabla^i h = \nabla^2 \nabla_j h -\nabla_j \nabla^2 h\,.
\end{equation}
In addition, one must employ \eqref{eq:effm} in order to verify that \eqref{eq:effp} remains invariant. Observe that the metric $\gamma_{ij}$ is the only gauge-invariant field in the effective theory. 

A particularly convenient viewpoint is that the symmetry relates effective theories that differ only by a shift of $s$. Namely, if the initial theory has a source $s$ and we transform the field variables with \eqref{eq:transmpfi} and
    \begin{align}
    % & m\to e^hm\,, \qquad p_i \to e^hp_i\,,\qquad  f_i\to f_i+\nabla_i h\,,\\
    & f_0\to f_0 -(\lambda^2+2\sigma)h-\nabla^2 h +\frac12 (\nabla s)^2-\frac12 (\nabla (s+h))^2\,,
\end{align}
then the resulting effective equations will take the same form, only now with $s$ replaced by $s+h$.

Although from the perspective of the complete gravitational system this is just a gauge transformation, when viewed as a symmetry of the membrane theory its key feature is that it mixes the non-dynamical source $s$ with the dynamical fields $m$ and $p_i$ that describe horizon deformations. As a result, $s$ can always be removed from the effective theory by taking $h=-s$, at the cost of changing the horizon profiles $m$ and $p_i$ (plus adding a potential flow to $f_i$ and shifting the gravitational potential $f_0$).

Conversely, any configuration in a theory with $s=0$ and a non-trivial bending profile $m(t,\vec{x})$ can be mapped via $h=-\ln m$ to a configuration in a different effective theory where the profile is uniform, $\bar m=1$, but the membrane possesses a compensating bending given by
\begin{equation}\label{eq:flatten}
    {\bar s}=-\ln m\,. 
\end{equation}
From the viewpoint of the full spacetime geometry, this transformation simply amounts to choosing Gaussian-normal coordinates adapted to the membrane, with the horizon at constant $\rho=0$. In the effective theory, however, it has a notable implication: the symmetry ensures that any spacetime-dependent profile $m(t,\vec{x})$ can be `flattened' into a uniform profile within another theory with an appropriate source $s(t,\vec{x})$. 

Alternatively, for a prescribed $m(t,\vec{x})$—possibly constant—the effective equations may be regarded as evolution equations for $s$, which then becomes dynamical. This perspective underlies the derivations of the stationary effective theory using different gauges in, e.g.,~\cite{Emparan:2015hwa,Suzuki:2015iha,Emparan:2016sjk}. We have not, however, investigated the properties of the resulting dynamical system of equations for $s$ in detail. 

Flattening the mass profile (or membrane height) with \eqref{eq:flatten} generally produces a non-vanishing momentum profile \eqref{eq:phtrans},
\begin{equation}
    \hat{p}_i\to \frac1{m}\left(\hat{p}_i-\nabla_i m\right)\,.
\end{equation}
However, if the original solution is a stationary one with $\nabla_i m =\hat{p}_i$, then the transformation yields
\begin{equation}\label{eq:trivconf}
    m\to 1\,,\qquad \hat{p}_i\to 0\,.
\end{equation}
That is, any stationary configuration can be transformed into a trivial configuration. We will illustrate this effect in Section~\ref{sec:examples}.

\paragraph{Equivalence-principle symmetry.} 
% The previous symmetry can also be used to add or remove a potential-flow component in the velocity source $f_i$. 
% Suppose that
% \begin{equation}\label{eq:potflow}
%     f_i =-\nabla_i \Psi(t,\vec{x})\,.
% \end{equation}
% Then the radial shift $\rho \to \rho+\Psi$ induces the transformations
% If we perform the transformations
% \begin{subequations}
%     \begin{align}
%  &  m\to e^{\Psi} m\,,\qquad p_i \to e^{\Psi} p_i\,,\qquad s\to s-\Psi\,,\\
%     &f_0 \to f_0 -(\lambda^2+2\sigma)\Psi-\nabla^2 \Psi +\frac12 (\nabla\Psi)^2-\nabla s\cdot \nabla\Psi+2\partial_t \Psi\,,
% \end{align}
% \end{subequations}
% then the effective equations preserve their form with the only change that the velocity source $f_i$ acquires a potential-flow component
% \begin{equation}
%     f_i\to f_i + \nabla_i \Psi\,.
% \end{equation}
% Therefore, any arbitrary potential flow can be removed from $f_i$ by a gauge transformation. The potential flow carries no vorticity, and its time dependence is reabsorbed into the transformed fields above.
The transformations \eqref{eq:fulltransf} can also be employed to add or remove a potential-flow component in the velocity source $f_i$, or to shift the gravitational potential $f_0$. 

This can also be achieved with the simpler symmetry transformation  
\begin{equation}
    f_i\to f_i +\nabla_i\Psi\,,\qquad f_0\to f_0 -\partial_t \Psi\,, 
\end{equation}
while leaving $m$, $\hat{p}_i$, $s$, and the metric unchanged. This shows that any $f_0$ can be removed by adding to $f_i$ a time-dependent potential flow such that $\partial_t \Psi=-f_0$. In other words, a gravitational force field can be replaced by a suitable acceleration field, which is a manifestation of the equivalence principle.

\paragraph{Scaling and Galilean symmetry.} 
A particular case of \eqref{eq:fulltransf} is the rescaling
\begin{equation}
    m\to \eta\, m\,,\qquad p_i \to \eta\,p_i\,,
\end{equation}
(so ${\hat p}_i \to \eta\, {\hat p}_i$) with constant $h=\ln \eta$. This allows us to arbitrarily choose a reference value for $m$.

Finally, in the absence of sources, the equations possess Galilean invariance under
\begin{align}
    x^i \to x^i - v^i t\,,\qquad p^i \to p^i + m v^i
\end{align}
with constant $v^i$ \cite{Emparan:2015gva}. In general, the sources will break this symmetry, but in some cases a boost-type symmetry may be inherited from the isometries of the background \cite{Licht:2022rke}. 

\paragraph{Gauge fixing vs.\ gauge freedom.} The discussion above shows that the effective equations \eqref{eq:effmp} are intrinsically gauge-dependent, since they are written in terms of quantities that transform non-trivially under \eqref{eq:rshift}. One may remove this dependence entirely by fixing the fields that vary in \eqref{eq:fulltransf}—for instance, by choosing $h$ so as to set $f_0=0$, or, more simply and naturally, by fixing
$s=0$ or imposing that $m$ be constant. In such gauges, the resulting equations yield a gauge-independent (though not manifestly covariant) formulation of the horizon dynamics.

However, the residual gauge freedom is not a mere redundancy. As in many areas of physics, different gauges highlight different aspects of the problem and may render specific classes of solutions considerably more tractable. For this reason, we prefer to keep the gauge freedom explicit within the effective theory, and in Section~\ref{sec:examples} we illustrate how working in different gauges can provide complementary and practically useful descriptions of the same physical configuration.

\subsection{Character of the equations} \label{subsec:remarks}

\paragraph{Non-linearity.} Observe that the equations are almost linear in $m$ and $\hat{p}_i$, except for the single term $\hat{p}_i \hat{p}_j/m$. Remarkably, this term alone encodes all of the non-linearity of Einstein's equations in the large-$D$ limit---and we recall that these effective equations are capable not only of reproducing black hole quasinormal modes, but also of describing fully non-linear horizon fluctuations, even including black hole collisions and mergers. We suspect that this striking simplicity in the limit $D\to\infty$ is not accidental, but a distilled reflection of a deeper mathematical structure underlying the full richness of Einstein’s theory.

\paragraph{Well-posedness of the dynamical evolution.}

The equations present the schematic form in \eqref{eq:dyneqs}, namely 
\begin{equation}\label{eq:parabU}
    \partial_t U-\nabla^2 U=\textrm{(lower-order in derivatives \& non-linear terms)}\,,
\end{equation}
with $U=(m,\hat{p}_i)$. The leading spatial operator on the left is the Laplacian acting diagonally on each component. Thus the highest-order spatial operator is elliptic in the spatial directions and the PDE is first order in time and second order in space---the hallmark of a parabolic system. Because the diffusion acts on all components (including the momentum components), the system is uniformly parabolic, which, with reasonable nonlinearities, implies well-posedness and smoothing of solutions. Therefore the system is, like Navier-Stokes-type systems, well posed as an initial-value problem. Indeed, \cite{Emparan:2016sjk} demonstrated that, for black branes (AF or AdS) the equations can be rewritten as a form of non-relativistic fluid dynamics. 

Although diffusive, the system can support propagating modes, not in the strict hyperbolic sense, since diffusion introduces infinite propagation (which is acceptable here, as the effective theory is non-relativistic) but in the form of damped, sound-like waves. A linearized analysis in the absence of sources shows that, depending on the sign of
\begin{equation}
    \lambda^2+2\sigma\gtrless 0\,,
\end{equation}
the system exhibits stable, damped sound modes (if negative), or unstable Gregory-Laflamme modes (if positive). The latter do not spoil well-posedness, since diffusion suppresses high-momentum modes; the instability only makes some long-wavelength modes exponentially grow, not an ill-posed Cauchy problem. Sources can enhance or hinder this instability, but the evolution of initial data remains well defined.

Summarizing, well-posedness of the initial value problem---existence, uniqueness, and continuous dependence on initial data---requires second-order spatial derivatives of $m$ and $\hat{p}_i$. This property is lost if one attempts to extend the range of $x^i$ from $O(1/\sqrt{D})$ to $O(D^0)$---a point to which we will return once more in Section~\ref{sec:O1eqs}.

\section{Particular instances and illustrative solutions}\label{sec:examples}

We next discuss several examples of effective equations, their properties, and their solutions, used to model black hole dynamics in different settings.\footnote{The corresponding backgrounds are described in Appendix~\ref{app:rescSigma}.} Afterwards, we show that stationary, homogeneous solutions exist only when the sources satisfy the $O(1/D)$ version of the soap-bubble condition.

\subsection{AF and AdS black branes}\label{subsec:eqsbbranes} 
We begin by setting all sources to zero,
\begin{equation}\label{eq:nosource}
    s=f=f_i=0\,,\qquad \gamma_{ij}=\delta_{ij}\,,
\end{equation}
so the effective equations we obtain are
\begin{subequations}\label{eq:bbranes}
    \begin{align}
    \partial_t m &=\partial^i\left(\partial_i m - p_i\right)\,,\label{eq:bbranesm}\\
    \partial_t p_i &= \left(\lambda^2+2\sigma\right) \partial_i m +\partial^j \left(\partial_j p_i -\frac{p_i p_j}{m} \right)\,.\label{eq:bbranesp}
\end{align}
\end{subequations}
As we saw, for both AF and AdS black branes we have $\lambda^2=1$, so setting $\sigma=0,\, -1$ respectively, we recover the effective equations governing the dynamics of these black branes, which were written in hydrodynamic form in \cite{Emparan:2016sjk}. The uniform black branes are the solutions with constant $m$ and zero $p_i$.

These equations have been widely studied, first to describe the Gregory–Laflamme instability of asymptotically flat uniform black branes \cite{Emparan:2015gva}, then to model localized black holes as `blobs' \cite{Andrade:2018nsz,Andrade:2019edf,Licht:2020odx,Suzuki:2020kpx,Licht:2022wmz}---we will discuss this next. In contrast, the AdS black brane is stable, and its effective theory has been applied to the dynamics of the holographic dual plasma \cite{Herzog:2016hob,Rozali:2017bll,Luna:2022tgh}.

These equations take the form \eqref{eq:parabU} of heat equations for $m$ and $p_i$, whose irreversible character reflects the dissipative nature of black hole horizons. This may seem puzzling, since at leading order in the large-$D$ expansion the entropy is proportional to the mass and hence conserved, making it unclear how the irreversible evolution should be understood. The resolution is that using $m$ and $p_i$ one can construct a quantity $S_1(t)$ which the equations of motion guarantee to be non-decreasing, $\partial_t S_1(t)\geq 0$ \cite{Andrade:2020ilm}. This quantity is precisely the $1/D$ correction to 
the black brane entropy obtained from the event horizon area. It would be interesting to extend this analysis to include sources.

\subsection{Schwarzschild as a blob on a black brane} \label{subsec:schblob}

It is known that the Schwarzschild black hole—and, in fact, also the Myers-Perry family—can be recovered as a Gaussian “blob” configuration of the asymptotically flat black string or black brane. Specifically, the equations \eqref{eq:mpeq} for a black string along $x$, namely,
\begin{subequations}\label{eq:breffeqs}
    \begin{align}
    \partial_t m &=\partial_x\left(\partial_x m - p_x\right)\,,\\
    \partial_t p_x &= \partial_x\left(\partial_x p_x+\partial_x m-\frac{p_x^2}{m}\right)\,,
\end{align}
\end{subequations}
admit the localized static solution
\begin{equation}\label{eq:blob}
    m(x)= e^{-x^2/2}\,,\qquad p(x)=-x e^{-x^2/2}\,,
\end{equation}
and, as shown in \cite{Andrade:2018nsz} (App.~A), the resulting spacetime metric
\begin{align}\label{eq:Schblob}
    ds^2 =& -\left(1- e^{-\rho-x^2/2}\right) dt^2 + \frac{2 dt}{n}\left(d\rho +x e^{-\rho-x^2/2}dx\right) \nn\\
    &+\left(1
    +\frac{x^2 e^{-\rho-x^2/2}}{n}\right)\frac{dx^2}{n} 
    +\left(1+\frac{2\rho}{n}\right)d\Omega_{n+1}\,,
\end{align}
corresponds precisely to the Schwarzschild black hole in the large-$D$ limit.

This construction, extended to rotating blobs on a black brane, forms the basis of the “blobology” approach to large-$D$ black holes---a highly versatile framework that has enabled the discovery of new black hole solutions and analytical calculations of quasinormal spectra for rotating black holes \cite{Andrade:2018nsz,Licht:2020odx,Suzuki:2020kpx,Licht:2022wmz}, and even the study of higher-dimensional black hole collisions, including scenarios violating cosmic censorship \cite{Andrade:2018yqu,Andrade:2019edf,Andrade:2020ilm}.

\subsection{Schwarzschild as forced black brane} \label{subsec:schlD}

The Schwarzschild black hole can also be described as a uniform black string (or black brane) solution of a different effective theory with non-trivial sources. This provides the simplest realization of the flattening transformation introduced in Section~\ref{subsec:membend}, so we examine it here in some detail.

To provide a useful additional perspective on the construction, we begin by recalling that in Section~\ref{sec:soapbub} we obtained the static Schwarzschild black hole as a spherical soap bubble embedded in Minkowski space. Following our discussion in Sec.~\ref{subsec:scaling}, we focus on a region of extent $\sim 1/\sqrt{n}$ around a maximal cycle of $S^{n+2}$, i.e., 
\begin{align}\label{eq:largenOmega}
    d\Omega_{n+2}&=d\theta^2+\cos^2\theta\, d\Omega_{n+1} \nn\\
    &\simeq \frac{dx^2}{n}+\left( 1-\frac{x^2}{n}\right)\, d\Omega_{n+1}\,,
\end{align}
where in the second line we set $\theta = x/\sqrt{n}$ and expanded for large $n$, keeping $x$ fixed. This a metric of the form of \eqref{eq:nearmax}. Then, using Eddington–Finkelstein coordinates and the radial coordinate $\rho$, the Minkowski metric becomes, in the leading large $n$ limit,
\begin{equation}\label{eq:largenMinks}
     ds^2 = -dt^2 + \frac{2}{n} dt\, d\rho +\frac{dx^2}{n} +\left(1+\frac{2\rho}{n}-\frac{x^2}{n}\right)d\Omega_{n+1}\,.
\end{equation}
The crucial element here is the term $-x^2/n$ in $g_{\Omega\Omega}$: from \eqref{eq:largenOmega} we see that it encodes the curvature along the direction $x$---the polar angle of the sphere $S^{n+2}$---of the membrane surface $M$ at constant $\rho$.\footnote{The term $2\rho/n$ does not represent new sources in the near-horizon geometry, since it is fixed by the Einstein equations, see \eqref{eq:sources} and \eqref{eq:nextsol}.} Without this term, the surface $M$ would instead extend along a flat, uniform direction $x$ in the background
\begin{equation}\label{eq:MinklD}
     ds^2 = -dt^2 + \frac{2}{n} dt\, d\rho +\frac{dx^2}{n} +\left(1+\frac{2\rho}{n}\right)d\Omega_{n+1}\,,
\end{equation}
and the membrane would describe a black string rather than a spherical black hole. That is, the embedding function $\phi(x)=1-x^2/n$ makes the membrane acquire spherical curvature. In Appendix~\ref{app:rescSigma} we extend this analysis to wider classes of metrics.

We can now write the effective equations for black holes in the Minkowski background \eqref{eq:largenMinks}. The sources are 
\begin{equation}
    \gamma_{xx}=1\,, \quad s(x)=-\frac{x^2}{2}\,,\quad f=f_i=0\,,
\end{equation}
with $\sigma=0\,,\kappa=+1\,,\lambda=1$. As we have argued, $s(x)$ enforces the spherical bending of the horizon. The effective equations become
\begin{subequations}\label{eq:scheffeqs}
    \begin{align}
    \partial_t m &=\left(\partial_x-x\right)\left(\partial_x m - p_x\right)\,,\\
    \partial_t p_x &= \left(\partial_x-x\right)\left(\partial_x p_x-\frac{p_x^2}{m}\right)-\partial_x m+p_x\,.
\end{align}
\end{subequations}
These admit the constant solution
\begin{equation}\label{eq:constm}
    m=1\,,\qquad  p_x=0\,, 
\end{equation}
and from \eqref{eq:nextsol} we find $G^{(1)}_{xx} = 2\rho$. Then, the complete solution for the metric is
\begin{equation}\label{eq:schlD}
    ds^2 = -\left(1-e^{-\rho}\right) dt^2 + \frac{2}{n} dt d\rho +\left(1+\frac{2\rho}{n}\right)\frac{dx^2}{n} +\left(1+\frac{2\rho}{n}-\frac{x^2}{n}\right)d\Omega_{n+1}\,.
\end{equation}
Thus, we recover precisely the large-$D$ limit of the Schwarzschild black hole \eqref{eq:schlarge}, with $d\Omega_{D-2}$ written as in \eqref{eq:largenOmega}.

This metric differs from \eqref{eq:Schblob} because each originates from a distinct effective theory of a black string---\eqref{eq:breffeqs} and \eqref{eq:scheffeqs}, with and without sources. A coordinate transformation mixing $\rho$ and $x$---namely, $\rho \to \rho + x^2/2$---allows the asymptotic source $s(x)$ to play the role of the non-trivial $m(x)$ profile in \eqref{eq:blob}.

The two theories describe the same black hole in strikingly different ways. In the blob picture \eqref{eq:blob}, the Gaussian profile appears because the uniform black string is Gregory-Laflamme-unstable, with the instability settling into a localized, stable black hole. By contrast, in \eqref{eq:constm} the source 
$s(x)$ stabilizes the uniform solution, suppressing all unstable quasinormal modes and removing the Gregory-Laflamme instability of the black string. Furthermore, while the blob picture easily allows one to study systems with multiple black holes, the theory \eqref{eq:scheffeqs} does not seem able to describe more than one---this would require modifying the source $s$ case by case.

It is straightforward to extend this construction to describe the Schwarzschild black hole as a uniformly extended black $p$-brane forced to bend into a sphere. One rewrites the Minkowski background as in \eqref{eq:bckgs} with the quadratic source \eqref{eq:gens} along $p$ coordinates, thus allowing the horizon to vary in multiple spatial directions---typically, two suffice to capture most black hole dynamics.

\subsection{AdS black strings}\label{subsec:adsbstring}
Let us now consider a negative cosmological constant, $\sigma=-1$, and take 
\begin{equation}
    \gamma_{xx}=1\,, \qquad s(x)=\frac{x^2}{2}\,,\qquad f=f_i=0\,.
\end{equation}
As before, we introduce a source $s(x)$, now with the opposite sign relative to the previous case; as shown in Appendix~\ref{app:rescSigma}, this is the appropriate choice for describing fluctuations of a black string in AdS. The effective equations become
\begin{align}
    \partial_t m &=\left(\partial_x+x\right)\left(\partial_x m - p_x\right)\,,\\
    \partial_t p_x &= \left(\partial_x+x\right)\left(\partial_x p_x-\frac{p_x^2}{m}\right)-p_x+%\left(1+\frac{\kappa}{r_0^2}\right)
    \lambda^2\partial_x m\,.
\end{align}
These equations have been derived in \cite{Emparan:2021ewh,Licht:2022rke} as the large-$D$ effective theory for AdS black strings (which come in three versions, depending on $\kappa$).

The constant solution $m=1$, $p=0$ corresponds to the uniform AdS black string. Refs.~\cite{Emparan:2021ewh,Licht:2022rke} further showed that the equations also admit a Gaussian blob solution, representing a Schwarzschild–AdS black hole. In a manner analogous to Section~\ref{subsec:schlD}, this black hole can alternatively be described as a uniform configuration in the AdS background \eqref{eq:bckgs}.

Including a Karch-Randall brane is straightforward, and it allows one to study the bulk holographic dual of black hole evaporation \cite{Emparan:2023dxm}.

\subsection{Further examples}

A slightly less general instance of the equations \eqref{eq:effmp} appears in \cite{Andrade:2018zeb}, who work with $\kappa=0$ and $\sigma=-1$ (so $\lambda=1$), in two spatial dimensions $x^i$, $i=1,2$, so that $R_{ij}=\gamma_{ij}R/2$, and with all sources switched on except for $s(t,x^i)$, which is gauge-fixed to 0. Ref.~\cite{Andrade:2018zeb} also employ a slightly different gauge for $A$, in which their function $\gamma_{tt}$ corresponds to $2f_0 + \mathrm{tr}(\partial_t \gamma)$ in our gauge.

In a different setting, the equations derived in \cite{Herzog:2017qwp} for a black hole in $\mathrm{AdS}_{D/2}\times S^{D/2}$ smeared over the $S^{D/2}$, correspond to $\sigma=-1$, $\kappa=+1$, with a source $s(\theta)=\cos\theta$ and $R_{\theta\theta}\sim 1/\sin^2\theta$. To obtain well-posed evolution, one must restrict to angles near $\theta=\pi/2$ of $O(1/\sqrt{n})$.

Although we do not study charged black holes here, we note that effective equations of this type have been obtained for charged systems, in different setups, in \cite{Emparan:2016sjk} and \cite{Iizuka:2018zgt}. 

Finally, we note that \cite{Jarvinen:2024vdi} develops a large-$D$-inspired framework for effective equations in strongly coupled holographic systems in finite dimensions. We expect that sources can be added following our general theory.

\subsection{Soap-bubble equation at $O(1/D)$}\label{subsec:soap1D}

Having identified the soap-bubble equation as the leading-order condition for stationary embeddings, we now show that the existence of uniform stationary solutions of \eqref{eq:effmp} requires the boundary sources in the near-horizon region to obey a subleading soap-bubble condition on scales $O(1/\sqrt{D})$.\footnote{We prove this within a fixed effective theory, without invoking the transformations in Section~\ref{subsec:membend} that relate different theories.}

To see this, we demand that the equations admit a solution where $m$ is uniform and constant (which we could set to one) and the velocity
\begin{equation}
    \hat{v}^i\equiv \frac{\hat{p_i}}{m} 
\end{equation}
is a time-independent Killing vector, i.e.,
\begin{equation}
    \nabla_{(i}\hat{v}_{j)}=\partial_t \hat{v}_j=0\,.
\end{equation}
This is well motivated, since expansion and shear lead to dissipation.\footnote{This argument, based on \cite{Caldarelli:2008mv}, will be reprised in Section~\ref{subsec:Msoap}.} We also assume that the sources $s$, $f_0$ and $\gamma_{ij}$ are time-independent and invariant under the action of $\hat{v}^i$, so that, in particular,
\begin{equation}
    \hat{v}^i\nabla_i s=0\,.
\end{equation}
The sources are nevertheless allowed to vary along directions not aligned with the velocity.

Under these conditions, the scalar equation \eqref{eq:effm} is automatically satisfied, while the vector equation  \eqref{eq:effp}, after applying identities like the Killing vector lemma $\nabla^2 \hat{v}_i=-R_{ij}\hat{v}^j$, becomes
\begin{equation}\label{eq:statvec}
    \nabla_j f_0+\frac{1}{2}\nabla_j \hat{v}^2-2\hat{v}^i\nabla_{[i}f_{j]}=0\,.
\end{equation}

Now we want to rewrite this effective equation using geometric invariants of the membrane, which lies at a constant-$\rho$ surface in the asymptotic region $\rho\to\infty$. The extrinsic curvature of such a surface can be computed for a metric in the class of \eqref{eq:BondiSachs} with metric functions \eqref{eq:sources}, \eqref{eq:nextsol} and \eqref{eq:nextAF}. The results we need are
\begin{align}
    K=\lambda D +O(D^0)\,,\qquad \sqrt{-g_{tt}}=\lambda +O(1/D)\,,
\end{align}
and
\begin{align}
   \nabla_i\left(\sqrt{-g_{tt}}K\right)= -\nabla_i \left(f_0+\text{tr}(\partial_t \gamma) +\partial_t s\right)+O(1/D)\,.
\end{align}
In the time-independent configurations we are considering, $\text{tr}(\partial_t \gamma)=\partial_t s=0$. We shall further assume that the velocity source $f_i$ is vorticity-free. While this is a less obvious condition to impose, it is not unreasonable, since vorticity in a horizon will lead to dissipation, if not at leading order in $1/D$, certainly once higher orders are included. Under this assumption, the last term in \eqref{eq:statvec} is absent. Notice that this is a condition on $f_i$ and not $p_i$, but it might be natural to also assume zero vorticity for the latter.

It is now straightforward to see that the resulting form of the vector effective equation
\begin{equation}\label{eq:presoap}
    \nabla_i \left(f_0+\frac{1}{2} \hat{v}^2\right)=0
\end{equation}
is, to first non-trivial order, equivalent to
\begin{equation}\label{eq:soap1D}
    \nabla_i\left(\sqrt{-g_{tt}(1-\boldsymbol{v}^2)}K\right)=0\,,
\end{equation}
once we account for the $1/\sqrt{D}$ scaling of lengths and include the leading redshift factor $\lambda$ at the membrane to identify the physical velocity as
\begin{equation}
    \boldsymbol{v}=\frac{\hat{v}}{\lambda\sqrt{D}}\,.
\end{equation}
Equation \eqref{eq:soap1D} is precisely the soap-bubble condition we set out to obtain. It is easy to verify that the quantity inside the brackets in \eqref{eq:presoap} is, up to a constant factor, the surface gravity of the horizon, so this equation amounts to the zeroth law of black holes.

It is also suggestive of a broader pattern. In Sections~\ref{subsec:schblob} and \ref{subsec:schlD} we saw that a uniform solution in a theory with sources can describe the same black hole as a non-uniform solution in a source-free theory—the two being related by a coordinate transformation that mixes radial and membrane directions, thereby trading sources for horizon profiles. Moreover, \cite{Emparan:2016sjk} showed that the source-free equations \eqref{eq:bbranes} can themselves be recast, for stationary configurations, as a soap-bubble condition at $O(1/D)$, with the membrane profile now encoded in $m$ and $p_i$. This suggests that, under suitably mild conditions on the sources, the effective equations \eqref{eq:effmp} should admit an analogous reformulation of the form \eqref{eq:soap1D}, with $m$ and $\hat p_i$ now determining the soap-bubble profile.

%Finally, observe that \eqref{eq:statvec} contains the static gravitational potential $f_0$ but not $s$. The latter should be viewed as the restriction of the membrane embedding to lengths $O(1/\sqrt{D})$ on $\Sigma_{n+1}^{(\kappa)}$. In the metric, $s$ does not appear multiplied by $\rho$, unlike $f_0$, which means it does not contribute to the surface gravity.

\section{Effective equations for $O(D^0)$ lengths and their ill-posedness}\label{sec:O1eqs}

The purpose of this section is to elaborate on the point we made in Section~\ref{subsec:scaling}, namely that equations with $O(D^0)$ range over the horizon are inadequate for dynamical evolution. We will also clarify why, despite this limitation, they remain valuable for other applications---most notably for extracting quasinormal modes.

To this end, we will derive effective equations of this type for the Schwarzschild black hole and discuss their main features.\footnote{These results were obtained with Kentaro Tanabe. Similar equations for rotating black holes have also been derived, but here we set the rotation to zero for simplicity.}

\subsection{Effective equations for the Schwarzschild black hole}

Take a metric ansatz
\begin{equation}\label{eq:O1ansatz}
    ds^2=r^2 G_{ij}(t,\vec{x},r)dx^i dx^j + r^2 \cos^2\theta\, d\Omega_{n+1}+2dt \left(dr-A(t,\vec{x},r)dt-F_i(t,\vec{x},r) dx^i\right)\,,
\end{equation}
where $x^i=(\theta,\phi)$, so we allow for fluctuations along a polar angle $\theta$ and motion along $\phi$, which we take to be a Killing direction to simplify the discussion. The radial coordinate near the horizon is chosen again as in \eqref{eq:nhrho}. We require the asymptotic behavior as $\rho\to\infty$
\begin{align}
    &A=\frac12+ O\left(n^{-1}\right)\,,\qquad F_i=O\left(n^{-1}\right)\,, \nn\\ 
    &G_{\theta\theta}=1+O\left(n^{-1}\right)\,,\quad G_{\phi\phi}=\sin^2\theta\left(1+ O\left(n^{-1}\right)\right)\,,\quad G_{\theta\phi}=O\left(n^{-1}\right)\,,
\end{align}
which gives an asymptotic Minkowski background,
\begin{equation}
    ds^2 \to r^2 d\Omega_{n+3} + 2 dt\, dr-dt^2 +O(1/n)\,,
\end{equation}
with $S^{n+3}$ written in the form
\begin{equation}
    d\Omega_{n+3}=d\theta^2 + \sin^2\theta\,d\phi^2 + \cos^2\theta\, d\Omega_{n+1}\,.
\end{equation}

This ansatz is very similar to \eqref{eq:BondiSachs}, but with the crucial difference that we have $G_{\theta\theta},\,G_{\phi\phi} =O(n^0)$, since we intend to allow fluctuations along the entire horizon $\theta\in [0,\pi/2]$. Less importantly, we also restrict to slow velocities $F_i\sim 1/n$, which helps to obtain the effective equations.

Solving the Einstein equations for \eqref{eq:O1ansatz} to leading order in $1/n$ determines
\begin{equation}
    2A=1-e^{-\rho}m(t,\theta)\,,\qquad F_i=\frac1{n} e^{-\rho}p_i(t,\theta)\,.
\end{equation}
When $m$ is constant and $p_i=0$ this reproduces the Schwarzschild solution \eqref{eq:schlarge}. Solving to the next order we obtain the effective equations for fluctuations around it,
\begin{subequations}\label{eq:scheffo1}
    \begin{align}
    \partial_t m+\tan\theta\,\partial_\theta m&=p_\theta \tan\theta\,,\label{eq:scheffa}\\
    \partial_t p_\theta +\tan\theta\, \partial_\theta p_\theta&=-\partial_\theta m +\left(1-\tan^2\theta\right)p_\theta +\tan\theta\,\frac{p_\theta^2}{m}\,,\label{eq:scheffb}\\
    \partial_t p_\phi +\tan\theta\, \partial_\theta p_\phi &= 2p_\phi +\tan\theta\, \frac{p_\phi p_\theta}{m}\,.\label{eq:scheffc}
\end{align}
\end{subequations}
Here we recognize many similarities with the effective equations \eqref{eq:effmp}, and more specifically, when $\theta$ is small and $p_\phi=0$, with the equations \eqref{eq:scheffeqs} for fluctuations of Schwarzschild. One difference is that \eqref{eq:scheffo1} become singular at $\theta=\pi/2$, but the key feature is that, unlike \eqref{eq:scheffeqs}, in \eqref{eq:scheffo1} there are no second-order spatial derivatives of $m$ and $p_i$. The ultimate reason for this is the degeneration of the principal symbol in the large-$n$ limit of the angular equation---that is, \eqref{eq:red1} vs.~\eqref{eq:red2}.

\subsection{Ill-posedness} 
As we discussed in Section~\ref{subsec:scaling}, second-order spatial derivatives are essential to obtain well-posed evolution, since they suppress the growth of high-momentum components in the initial data.

To see how this problem arises in \eqref{eq:scheffo1}, we set for simplicity $p_\phi=0$ (the argument does not change otherwise). Noting that we can use \eqref{eq:scheffa} to eliminate $p_\theta$, we introduce a new coordinate ${\hat t}$ by
\begin{equation}
    \partial_{\hat t}=\partial_t +\tan\theta\, \partial_\theta\,,
\end{equation}
to find that \eqref{eq:scheffb} takes the form
\begin{equation}\label{eq:scheffd}
    \partial_{\hat t}^2 m+\tan\theta\,\partial_\theta m = \textrm{(lower order terms)}\,.
\end{equation}
This is like \eqref{eq:parabol} and therefore defines an ill-posed `reversed diffusion problem' with the uncontrolled growth of small-scale initial fluctuations in \eqref{eq:parabol2}. 

We can also reach this conclusion with a more conventional analysis of the principal symbol of \eqref{eq:scheffo1}, again with $p_\phi=0$. We write the equations as
\begin{equation}\label{eq:principal}
    \partial_t U + B(\theta)\,\partial_\theta U=\textrm{(lower-order terms)}\,,
\end{equation}
where $U=(m,p_\theta)^T$ and
\begin{equation}
   B=
  \begin{pmatrix}
    \tan\theta & 0 \\
    1 & \tan\theta  
  \end{pmatrix}\,.
\end{equation}
Being triangular with degenerate eigenvalues, this matrix cannot be diagonalized, so the system is weakly hyperbolic, but not strongly hyperbolic (including \eqref{eq:scheffc} does not modify this conclusion).

To understand why this kills well-posedness, we freeze the coefficients, $\tan\theta\to c$, and study the linear system
\begin{equation}
    \partial_t U + B\,\partial_\theta U=0\,,
\end{equation}
with 
\begin{equation}
B=
  \begin{pmatrix}
    c & 0 \\
    1 & c  
  \end{pmatrix} = cI + N\,,\quad \textrm{where}~N=\begin{pmatrix}
    0 & 0 \\
    1 & 0  
  \end{pmatrix}\,,
\quad N^2=0\,.
\end{equation}
Fourier transforming $\partial_\theta \to ik$ gives
\begin{equation}
    \partial_t U_k =-ik B\, U_k\,,
\end{equation}
which is solved by
\begin{equation}\label{eq:Usol}
    U_k(t) =e^{-ik B t}U_k(0)= e^{-ik ct}(I-ik N t)U_k(0)\,.
\end{equation}
The factor $I-ik Nt$ produces an amplitude growing like $|k|t$. That is, if we take initial data concentrated at large wavenumber $k$, then
\begin{equation}
    \left\Vert U_k(t)\right\Vert \sim (1+ |k| t)|U_k(0)|\,,
\end{equation}
so for $k\to\infty$ this unbounded factor destroys continuous dependence---small initial data with arbitrarily large momentum content can produce $O(1)$ outputs in arbitrarily small time. Although the solution \eqref{eq:Usol} differs from \eqref{eq:parabol2}---which arises after reducing the equations to \eqref{eq:scheffd} and setting the right-hand side to zero, i.e., a different truncation---the main conclusion remains the same.

A typical cure is to introduce a small viscous term $\sim\partial_\theta^2$ that suppresses the unruly short-wavelength oscillations, yielding a well-posed parabolic (or hyperbolic-parabolic) problem. In our system, this behavior is naturally restored by the restriction to small angles $\theta = x/\sqrt{n}$, c.f.~\eqref{eq:scheffeqs}, which is the regime where the relevant large-$D$ dynamics of the black hole takes place.

\subsection{Stationary solutions and quasinormal modes}

Although ill-suited for dynamical evolution, equations of the type of \eqref{eq:scheffo1} may be sensibly used for the study of stationary configurations\footnote{The extension of \eqref{eq:scheffo1} to rotating black holes can be used to obtain stationary `bumpy black holes' \cite{Suzuki:2015iha} and black rings \cite{Tanabe:2015hda}.} and linearized perturbations, i.e., quasinormal modes. It is illuminating to examine how the latter are obtained. Set
\begin{equation}
    m(t,\theta)=1+e^{-i\omega t}\mu(\theta)\,,\qquad p_i(t,\theta)=e^{-i\omega t}\pi_i(\theta)\,,
\end{equation}
and expand the equations to linear order in $\mu$ and $\pi_i$. Using \eqref{eq:scheffa} to eliminate $\pi_\theta$ in \eqref{eq:scheffb}, we obtain an equation whose general solution is
\begin{equation}
    \mu(\theta)= c_1 \left(\sin\theta\right)^{1/2+i\omega + a}+c_2 \left(\sin\theta\right)^{1/2+i\omega - a}\,,
\end{equation}
with 
\begin{equation}
    a=i\sqrt{i\omega-\frac14}\,.
\end{equation}
If we require that $\mu$ and its derivatives are regular at $\theta=0$, then we must have
\begin{equation}\label{eq:regular}
    1/2+i\omega \pm a =\ell\,,
\end{equation}
with $\ell$ a non-negative integer. Solving for $\omega$ gives the quasinormal mode frequencies
\begin{equation}\label{eq:scqnm}
    \omega =\pm \sqrt{\ell-1} -i(\ell-1)\,.
\end{equation}
In a similar way, \eqref{eq:scheffc} yields another quasinormal mode frequency,
\begin{equation}\label{eq:vqnm}
    \omega = -i(\ell-2)\,,
\end{equation}
with $\pi_\phi(\theta)=(\sin\theta)^{\ell}$. These are the correct frequencies of the scalar and vector quasinormal modes of the Schwarzschild black hole in the large-$D$ limit, first obtained in \cite{Emparan:2014aba}, later derived from the M equations in \cite{Bhattacharyya:2015dva}, and reproduced using the `blob' picture in \cite{Andrade:2018nsz}.

Although the equations \eqref{eq:scheffo1} reproduce the required quasinormal frequencies, this does not imply they capture genuine time-dependent dynamics, since an important subtlety underlies the derivation. The quantization condition \eqref{eq:regular} does not come from boundary conditions at $\theta\in[0,\pi/2]$---the equations actually diverge at $\theta=\pi/2$---but from imposing `initial-value’ conditions at $\theta=0$. This reflects the reversed-parabolic nature of \eqref{eq:scheffd}: it evolves along $\theta$ rather than physical time $t$. Although this procedure pleasingly yields the correct quasinormal spectrum, the reversed-parabolic nature of \eqref{eq:scheffd} precludes a well-posed dynamical evolution in physical time.

\section{Effective equations from the M-formalism}\label{sec:HfromM}

A central theme of this work is that the H and M approaches, despite their different formulations, encode the same underlying physics of large-$D$ black holes. To make this connection explicit, and to clarify how the two viewpoints organize the same information with different degrees of manifest covariance, we now show how the soap-bubble equations and the well-posed dynamical equations \eqref{eq:effmp} follow from the covariant M-formalism of \cite{Bhattacharyya:2015dva,Bhattacharyya:2015fdk}. Equations on $O(D^0)$ scales, like those in Section~\ref{sec:O1eqs}, can likely be derived similarly, but are expected to remain ill-posed.

\subsection{Setup and equations}

For completeness and to fix notation, let us briefly recall the M formalism of \cite{Bhattacharyya:2015dva,Bhattacharyya:2015fdk}.
It begins with an ansatz for geometries built over a background spacetime---Minkowski or (A)dS---of the form
\begin{equation}\label{eq:ksform}
    ds^2=ds^2_\textrm{(bkg)}+\left(\frac{r_0}{r}\right)^{D-3}(n_M-u_M)(n_N-u_N)dx^M dx^N\,.
\end{equation}
The formalism is designed to preserve manifest covariance, so the coordinates $x^M$ comprise both the $p+1$ directions tangent to the would-be membrane and the $n+1$ directions of the constant-curvature space previously denoted $\Sigma_{n+1}^{(\kappa)}$. A main goal is to treat all these directions on equal footing without distinguishing a preferred membrane subset. 
\emph{This is the key difference from the H approach}, whose effective equations explicitly single out the `active' membrane directions.\footnote{However, the soap-bubble equation is fully covariant.} As a result, the M formalism is better suited for analyzing general, covariant properties of the membrane as an object embedded in the background spacetime, whereas the H formalism is more effective for explicitly studying the dynamics of the horizon.

If we take $n_M = \partial_M r$ as the normal vector, and a constant velocity $u_M$ orthogonal to it, unit-normalized with respect to the background metric, the geometry describes a boosted black hole.\footnote{With the radius function $r^2(x)=x^M x^N (g_{MN}^\textrm{(bkg)}+u_M u_N)$.} Although $n$ has unit norm in the background metric, in the full spacetime we have
\begin{align}
    g^{MN} n_M n_N &=1 -\left(\frac{r_0}{r}\right)^{D-3}\,,
\end{align}
so the horizon is a null hypersurface generated by $u$ at $r=r_0$. The membrane picture emerges by viewing this hypersurface as a submanifold of the background: the membrane worldvolume (codimension-one and timelike) with unit normal $n_M$ and a tangent velocity vector $u_M$.

To introduce fluctuations, one makes $r$ and $u$ functions of $x^M$. %, while $n_M$ becomes the normal to the fluctuating membrane and $u_M$ remains tangent to it, $ g_\textrm{(bkg)}{}^{MN} n_M u_N=0$.
For the membrane, this means that it is now specified by the embedding function $r(x^M)$ and the local velocity vector $u(x^M)$. The shape of the membrane in the background is covariantly captured by the extrinsic curvature tensor
\begin{equation}
    K_{MN} = (\delta_M^P-n_M n^P) \nabla_P n_N\,, 
\end{equation}
We denote its pull-back on the membrane worldvolume by $\mathcal{K}_{\mu\nu}$ (with trace $\mathcal{K}=K$), and henceforth use Greek indices for such pull-backs.

When $r$ and $u$ vary with the coordinates, the metric \eqref{eq:ksform} no longer solves Einstein’s equations exactly. This is where the large-$D$ expansion enters: by assuming that, as $D\to\infty$, the membrane fluctuates along only a finite number of directions, one can systematically construct a solution by adding corrections order by order in $1/D$. These corrections are confined within a distance $\sim O(1/D)$ from the horizon.

Conceptually, this procedure parallels the H approach: the radial dependence of Einstein’s equations can be integrated at each order in $1/D$, and one obtains a set of constraint equations involving only the fields defined on the membrane, namely the velocity $u_\mu$ and the extrinsic curvature $\mathcal{K}_{\mu\nu}$. These constraints constitute the effective equations, which take the form
\begin{subequations}\label{eq:Meqs}
    \begin{align}\label{eq:scalarM}
        \hat{\nabla}_\mu u^\mu = 0\,,
    \end{align}
\begin{align}\label{eq:vectorM}
    \left(\frac{\hat{\nabla}^2u_\nu}{\mathcal{K}}-\frac{\hat{\nabla}_\nu \mathcal{K}}{\mathcal{K}}+u^\alpha\mathcal{K}_{\alpha\nu}-u^\alpha\hat{\nabla}_\alpha u_\nu\right)\mathcal{P}_\mu^\nu = 0\,,
\end{align}
\end{subequations}
where $\hat{\nabla}$ is the covariant derivative for the metric induced on the membrane, and 
\begin{equation}
    \mathcal{P}_\mu^\nu=\delta_\mu^\nu +u_\mu u^\nu
\end{equation}
is the spatial projector orthogonal to the membrane velocity.

Although these equations do not explicitly refer to a near-horizon region or to preferred membrane directions, these notions enter in their derivation and are required when applying the formalism to any concrete black hole. One introduces the large-$D$ near-horizon radial coordinate
\begin{equation}\label{eq:rrho}
    r=r_0\left(1+\frac{\rho}{D}\right)\,,
\end{equation}
which coincides with \eqref{eq:nhrho} to leading order, keeping the horizon scale $r_0$ explicit, and then specifies the membrane by
\begin{equation}\label{eq:rhoy}
    \rho=y(x^\mu)\,,
\end{equation}
where $x^\mu$, $\mu=1,\dots, p+1$ denote the fluctuation coordinates---the ones that in the H approach are usually referred to as the membrane directions.
Then the normal is
\begin{equation}
    n = \frac{d\rho-\partial_\mu y\, dx^\mu}{|d\rho-\partial_\mu y\, dx^\mu|}\,,
\end{equation}
with $u_M$ chosen to be orthogonal to it, and 
\begin{equation}
    \mathcal{K}_{\mu\nu}=K_{\mu\nu}+\partial_\mu y\, K_{\rho\nu}+\partial_\nu y \, K_{\mu\rho}+\partial_\mu y\,\partial_\nu y\, K_{\rho\rho}\,.
\end{equation}

\subsection{Soap-bubble equation}\label{subsec:Msoap}

We now reproduce the argument in \cite{Dandekar:2017aiv} that the covariant equations \eqref{eq:Meqs} imply the soap-bubble condition \eqref{eq:soapbubble} for stationary configurations.

In \cite{Bhattacharyya:2016nhn,Dandekar:2017aiv} it was shown that \eqref{eq:Meqs} are equivalent to the conservation of the stress tensor
\begin{equation}
    16\pi T_{\mu\nu}=\mathcal{K}\mathcal{P}_{\mu\nu}-2\sigma_{\mu\nu}+\mathcal{K}_{\mu\nu}-\mathcal{K}g_{\mu\nu}\,,
\end{equation}
where $\sigma_{\mu\nu}$ is the shear of the velocity field $u_\mu$. The last two terms are the Brown-York stress tensor for the membrane hypersurface, which is identically conserved and therefore will not play a role here.

In a stationary configuration, the shear must vanish, as any nonzero shear would produce dissipation and increase entropy. As shown in \cite{Caldarelli:2008mv}, imposing in addition that the divergence vanishes, i.e., \eqref{eq:scalarM}, requires the velocity to be aligned with a timelike Killing vector field $k$ on the membrane,
\begin{equation}
    u^\mu =\gamma\, k^\mu\,,\qquad \gamma=(-k\cdot k)^{-1/2}\,.
\end{equation}
This condition also implies that
\begin{equation}
    {\hat\nabla}_\nu\mathcal{P}^\nu_\mu=u^\nu{\hat \nabla}_\nu u_\mu=-{\hat \nabla}_\mu \ln \gamma\,.
\end{equation}
It then follows that
\begin{align}
    16\pi{\hat \nabla}_\nu T^\nu_\mu&={\hat\nabla}_\nu\left(\mathcal{K}\mathcal{P}^\nu_\mu\right)\nn\\
    &=\mathcal{P}^\nu_\mu\,{\hat\nabla}_\nu\mathcal{K}
    -\mathcal{K}\,{\hat \nabla}_\mu \ln \gamma \nn\\
    &=\gamma\,\mathcal{P}^\nu_\mu\hat{\nabla}_\nu\left(\gamma^{-1}\mathcal{K}\right)\,,
\end{align}
so stress-energy conservation directly leads to \eqref{eq:soapbubble}, with the integration constant identified as twice the black hole surface gravity.

\subsection{Membrane dynamics in flat space and AdS}
Before presenting the full derivation of the H-type equations from the M-formalism, it is useful to illustrate the main ideas in the simplest case: the black-brane equations \eqref{eq:bbranes}, whose derivation starting from \eqref{eq:Meqs} was given in \cite{Dandekar:2016jrp}.

Consider Minkowski or AdS written in the large $D$ limit as
\begin{equation}
    ds^2=-\lambda^2 dt^2+\frac{2}{D}dtd\rho+%\left(1+\sigma\frac{2\rho}{D}\right)
    \frac{\delta_{ij}}{D}dx^i dx^j + r_0^2\left(1+\frac{2\rho}{D}\right)d\Sigma^{(\kappa)}_{n+1}\,,
\end{equation}
(see Appendix~\ref{app:rescSigma}). Even though in these two cases we have $\lambda=1$, we will retain it explicitly for later reference.
The metric induced  on the membrane defined by \eqref{eq:rhoy} is
\begin{equation}\label{eq:indmem}
    \left.ds^2\right|_M=- \lambda^2 dt^2+\frac{2\partial_i y}{D}dtdx^i+\frac{\delta_{ij}}{D}dx^i dx^j + r_0^2\left(1+\frac{2y}{D}\right)d\Sigma^{(\kappa)}_{n+1}\,.
\end{equation} 

Using this metric to normalize the velocity $u$ requires that
\begin{align}
    & u^\mu=\frac{1}{\lambda}\left(1,v^i\right)+O(1/D)\,,
\end{align}
 where $v^i$ acts as a Galilean velocity. 
The scalar equation \eqref{eq:scalarM} then becomes
\begin{equation}\label{eq:scalarM2}
    \nabla_i  v^i+\dot{y}+v^i(\nabla_i y)=0\,.
\end{equation}
We must now translate from $(y,v^i)$ to the variables $(m,p^i)$ of the H-formalism. The relation is straightforward,
\begin{align}\label{eq:effmap}
    \ln y(t,x^i)=m(t,x^i)\,,\qquad v^j(t,x^i)=\frac{p^j(t,x^i)-\nabla^jm(t,x^i)}{m(t,x^i)}\,,
\end{align}
which is valid even when the metric along $x^i$ is not flat.
Then \eqref{eq:scalarM2} becomes
\begin{equation}
    \partial_t m=\partial^i\partial_im-\partial_i p^i\,,
\end{equation}
which is the first of the H-equations, \eqref{eq:bbranesm}.

Next, the terms entering the vector equation \eqref{eq:vectorM} can be computed for the membrane \eqref{eq:indmem}, to find
\begin{subequations}\label{eq:vectorterms}
\begin{align}
    & \frac{\hat{\nabla^2}u_i}{\mathcal{K}}=\frac{\partial^2 v_i}{\lambda^2 D}+O(1/D^2)\,,\\
    & \frac{\hat{\nabla}_i\mathcal{K}}{\mathcal{K}}=\frac{1}{\lambda^2 D}\left(-\partial^j\partial_j\partial_i y-\partial_i \partial^2 y-\frac{\kappa}{r_0^2}\partial_i y\right)+O(1/D^2)\,,\\
    & u^\mu \mathcal{K}_{\mu i}=\frac{1}{\lambda^2 D}\left(-\partial_i \dot{y}-\sigma (v_i+\partial_i y)-v^j\partial_j\partial_i y\right)+O(1/D^2)\,,\\
    & u^\mu \hat{\nabla}_\mu u_i=\frac{1}{\lambda^2 D}\left(\dot{v}_i-\sigma \partial_i y+v^j\partial_j v_i\right)+O(1/D^2)\,,\\
    & \mathcal{P}^j_i=\delta^j_i+O(1/D)\,,\\
    & \mathcal{P}^t_i \left(\frac{\hat{\nabla^2}u_t}{\mathcal{K}}-\frac{\hat{\nabla}_t\mathcal{K}}{\mathcal{K}}+u^\mu \mathcal{K}_{\mu t}-u^\mu \hat{\nabla}_\mu u_t\right)=\frac{\sigma}{\lambda^2 D}\left(v_i+\partial_i y\right)+O(1/D^2)\,.
\end{align}
\end{subequations}
Putting these all together into \eqref{eq:vectorM}, we obtain
\begin{align}
    \partial_t{p}_i=\partial^2 p_i-\partial^j\left(\frac{p_jp_i}{m}\right)+\left(\lambda^2 +2\sigma\right)\partial_i m\,,
\end{align}
which is indeed the vector H-equation \eqref{eq:bbranesp}.

\subsection{General equations}

We now extend the analysis by applying the covariant equations \eqref{eq:Meqs} to the general metrics \eqref{eq:BondiSachs} and \eqref{eq:sources}, thereby deriving the general effective equations \eqref{eq:effmp}. 

In this case, the metric induced on the membrane at leading order is
\begin{equation}
    ds^2=-\lambda^2dt^2+\frac{2}{D}\left(\partial_i y-f_i\right)dtdx^i+\frac{\gamma_{ij}}{D}dx^idx^j+r_0^2\left(1+\frac{2}{D}(y+s)\right)d\Omega_{n+1}.
\end{equation}
Observe that the source $f_0$ does not appear here. As discussed in Section~\ref{subsec:membend}, $f_0$ has a different status from the other sources: it is not the background value at infinity of a metric coefficient, but instead enters through a correction term that grows toward the near-horizon boundary as $\rho/D$. Hence, as we argued, it is directly related to the local surface gravity at $O(1/D)$, up to a local boost.

\paragraph{Scalar equation.}
The scalar equation is the incompressibility of the velocity field \eqref{eq:scalarM}.
To write it for the metrics \eqref{eq:BondiSachs} and \eqref{eq:sources} we need several geometric magnitudes at the relevant orders in a $1/D$ expansion, which can be found in Appendix~\ref{app:techdet}. After some manipulation, we find that \eqref{eq:scalarM} takes the form
\begin{equation}\label{eq:scaleq0}
    \nabla_i  v^i+\frac{1}{2}\textrm{tr} (\dot{\gamma})+\dot{y}+\dot{s}+v^i(\nabla_i y + \nabla_i s)=0\,.
\end{equation}
Now $\nabla$ (without a hat) denotes the covariant derivative associated with the spatial membrane metric $\gamma_{ij}$.
Note that this equation is no longer Lorentz invariant, as the introduction of a non-relativistic velocity field breaks relativistic symmetry.
After substituting \eqref{eq:effmap} into \eqref{eq:scaleq0}, we easily recover \eqref{eq:effm}.

\paragraph{Vector equation.}
Finally, we obtain the form that \eqref{eq:vectorM} takes for these metrics. Again, lengthy details are postponed to Appendix~\ref{app:techdet}. For the different terms that enter into \eqref{eq:vectorM}, we find
\begin{subequations}\label{eq:vecterms}
\begin{align}
    \frac{\hat{\nabla}^2 u_j}{\mathcal{K}}=&\frac{1}{\lambda^2 D}\left[\left(\nabla^i+\nabla^i y+\nabla^i s\right)\left(\nabla_i v_j+\nabla_{[j}f_{i]}+\frac{1}{2}\dot{\gamma}_{ij}\right)\right]+O(1/D^2)\,\\
    \frac{\hat{\nabla}_j \mathcal{K}}{\mathcal{K}}=&\frac{1}{\lambda^2 D}\Bigg[-\left(\nabla^i y+\nabla^i s\right)\nabla_i\nabla_j y-\nabla_j\nabla^2 y -\lambda^2\nabla_j y-\nabla^i y\nabla_i\nabla_j s \nn\\
    &+\nabla_j\left(s_a^{(1)}+A^{(1)}+\dot{s}+f\cdot\nabla s+\frac{1}{2}f^2+R-\nabla^2 s\right)\Bigg]+O(1/D)  \,\\  
    u^\mu \mathcal{K}_{\mu j} = &\frac{1}{\lambda^2 D}\Bigg[-\nabla_j \dot{y}+\left(\nabla^i y - f^i\right)\left(\frac{1}{2}\dot{\gamma}_{ij}+\nabla_{[i}f_{j]}\right) +\nabla_j A^{(1)} \nn\\ &-\frac{\partial_\rho F^{(1)}_j}{2}
    +v^i\left(-\nabla_i\nabla_j (y+s)+R_{ij}\right)-\sigma v_j\Bigg]+O(1/D^2)\,\\
    u^\mu \hat{\nabla}_\mu u_j =&\frac{1}{\lambda^2 D}\left[\dot{v}_j-\dot{f}_j+\nabla_j A^{(1)}+v^i\left(\nabla_i v_j+2\nabla_{[j}f_{i]}\right)\right]+O(1/D^2)\,\\
    \mathcal{P}^t_i u^\mu \mathcal{K}_{\mu t} =& \frac{\sigma}{\lambda^2 D}\left(v_i+\nabla_i y -f_i\right)+O(1/D^2)\,.
\end{align}
\end{subequations}
These generalize \eqref{eq:vectorterms}. Putting them together and transforming variables as in  \eqref{eq:effmap} (see details in Appendix~\ref{app:veceqn}), we reproduce the vector equation of the H-formalism \eqref{eq:effp}.

\section{Conclusions and Outlook}\label{sec:conclusion}

Our main result is the general effective equations \eqref{eq:effmp} governing black hole dynamics in the large-$D$ expansion with well-posed evolution. These equations are part of a universal structure: stationary configurations satisfy the simple geometric embedding condition \eqref{eq:soapbubble}, while dynamical fluctuations of amplitude $O(1/D)$ and range $O(1/\sqrt{D})$ obey the nonlinear, forced evolution equations \eqref{eq:effmp}, valid for generic backgrounds and without symmetry restrictions beyond a cyclic factor of dimension $n\gg 1$.

We have shown that these equations arise from the near-horizon expansion in Bondi–Sachs gauge \eqref{eq:BondiSachs}, where, at large $D$, the metric functions \eqref{eq:sources} depend on the dynamical variables $m(t,x^i)$, $p_i(t,x^j)$ and the tidal sources $s$, $f_0$, $f_i$, and $\gamma_{ij}$, which encode the influence of the surrounding spacetime. This framework retains a built-in flexibility stemming from a residual radial gauge freedom, allowing the membrane bending to be represented either through the profile of $s$ or through that of $m$. It also makes transparent the interchangeability between an acceleration field and a gravitational force, explicitly realizing the equivalence principle.

While constrained by the soap-bubble condition, the sources can also model independent external forces. When the near-horizon region is treated as an autonomous system, the sources can, in principle, be chosen arbitrarily; however, a proper coupling to a background would fix them (up to gauge) through a matched-asymptotics construction. Exploring this further could provide a better understanding of how time-dependent membrane dynamics are captured within the $1/D$ expansion.

The leading large-$D$ effective theory that we have presented is naturally limited to horizon fluctuations of amplitude $O(1/D)$ and extent $O(1/\sqrt{D})$. The former reflects the localization of dynamics near the horizon, while the latter is less obvious but is necessary for well-posed evolution. Although these may seem to impose significant limitations in the reach of the large-$D$ description, they are simply a reflection that the most important dynamics of the horizon occur within these ranges. An interesting open direction is to derive the general H-type equations for $O(D^0)$-range horizon fluctuations, investigate their connection to the soap-bubble condition, and prove their ill-posedness. 

We have also clarified the connection to the covariant M formulation \cite{Bhattacharyya:2015dva,Bhattacharyya:2015fdk}---a correspondence that we expect to hold quite generally, at least when requiring well-posed evolution. The M approach provides a compact and elegant framework that appears to encompass all H-type equations---a genuinely powerful feature. Its covariance is conceptually appealing, although it can make some structural aspects less transparent—most notably, the distinction between stationary embeddings and near-horizon dynamical evolution, and the limitations of the range of dynamics captured by the expansion.
We expect that the connections we have drawn clarify the circumstances in which each framework is best applied.

Natural future directions within the H formalism include incorporating sources for charged black holes \cite{Emparan:2016sjk,Andrade:2018rcx}, deriving a general entropy current and second law \cite{Andrade:2020ilm}, and systematically including next-order $1/D$ and higher-curvature corrections.

In conclusion, the simplicity and universality of the large-$D$ effective equations, together with a better understanding of their structure, scope, and limitations, make them a powerful and practical tool for addressing black hole problems that may otherwise be intractable.

\section*{Acknowledgments}

We are grateful to Sayantani Bhattacharyya, Shiraz Minwalla and Arunabha Saha for discussions. We are also deeply indebted to Keisuke Izumi, Ryotaku Suzuki, and Kentaro Tanabe for initial collaboration along related directions, and for sharing their many insights into the large $D$ program. This work has been supported by MICINN grant PID2022-136224NB-C22, AGAUR grant 2021 SGR 00872, and the ``Unit of Excellence María de Maeztu'' grant CEX2024-001451-M funded by MICIU/AEI/10.13039/501100011033.

\appendix

\section{General equations in Bondi-Sachs gauge}\label{app:BondiSachseqns}

Here we derive the Einstein equations for arbitrary $D$ (and $n$) in the ingoing Bondi-Sachs metric ansatz \eqref{eq:BondiSachs}, drawing from the analysis in \cite{Chesler:2010bi}.

Note first that there is a residual shift symmetry that leaves the ansatz invariant
\begin{equation}
r \to r + \delta\lambda(x).
\end{equation}
Because of this shift symmetry, it is natural to define the following derivatives
\begin{subequations}
\begin{align}
d_+ &\equiv \partial_t + A(X)\, \partial_r, \label{eq:dplus} \\
d_i &\equiv \partial_i + F_i(X)\, \partial_r. \label{eq:di}
\end{align}
\end{subequations}

Diffeomorphisms of the $p$-dimensional spatial (non-$\Sigma$ and non-radial) directions transform the
various metric functions covariantly in the usual way. We will therefore view $G_{ij}$ as a $p$-dimensional
spatial metric and use it to raise and lower latin indices. However, we should be careful with its
dependence on $t$ and $r$ because temporal and radial differentiation do not commute with the raising
and lowering of indices. To handle this, we define separate tensors that give $t$ and $r$ derivatives
with lower indices. That is, we define the tensor $G'_{ij} \equiv \partial_r G_{ij}$, and then raise and lower its indices
with the metric, so $G'^{ij} = G^{ik}G^{jl}(\partial_r G_{kl})$. With this definition, the derivative of the inverse metric
can be written
\begin{equation}
\partial_r G^{ij} = -G^{ik} G^{jl} \partial_r G_{kl} = -G'^{ij}.
\end{equation}
We make similar definitions for $G''_{ij}$, $F'_j$, and also for $d_+ G_{ij}$, etc.

We can create a modified Christoffel connection by replacing $\partial_i$ with $d_i$ in the usual expressions:
\begin{align}
\tilde{\Gamma}^i_{jk} & \equiv \frac{1}{2} G^{il} 
\left( d_k G_{lj} + d_j G_{lk} - d_l G_{jk} \right)\nn\\
&= \frac{1}{2} G^{il} \big( G_{lj,k} + G_{lk,j} - G_{jk,l} + G'_{lj} F_k + G'_{lk} F_j - G'_{jk} F_l \big).
\end{align}
This connection is torsion free: $\tilde{\Gamma}^i_{jk} = \tilde{\Gamma}^i_{kj}$.
We can analogously define a modified covariant derivative $\tilde{\nabla}$ in the same way, so for example
\begin{equation}
\tilde{\nabla}_i V_j = d_i V_j - \tilde{\Gamma}^k_{ij} V_k.
\end{equation}
The modified covariant derivative is metric-compatible: $\tilde{\nabla}_i G_{jk} = 0$.

Similarly, we can define the modified Riemann tensor
\begin{equation}
\tilde{R}^i{}_{jkl} \equiv d_k \tilde{\Gamma}^i_{jl} - d_l \tilde{\Gamma}^i_{jk} 
+ \tilde{\Gamma}^i_{mk} \tilde{\Gamma}^m_{jl} - \tilde{\Gamma}^i_{ml} \tilde{\Gamma}^m_{jk},
\end{equation}
as well as the Ricci tensor and scalar 
\begin{equation}
\tilde{R}_{jk} \equiv \tilde{R}^i{}_{jik}, \qquad \tilde{R} \equiv \tilde{R}^k{}_k.
\end{equation}
Note, though, that the modified Riemann tensor does not have the same index symmetries as the usual Riemann tensor.
Likewise for the Ricci tensor (i.e., the modified Ricci tensor is not symmetric).

We also define the “magnetic” field strength two-form
\begin{equation}
\Omega_{ij} \equiv \tilde{\nabla}_i F_j - \tilde{\nabla}_j F_i 
= \partial_i F_j - \partial_j F_i + F_i F'_j + F_j F'_i,
\end{equation}
and the “electric” one-form
\begin{equation}
E_i \equiv d_+ F_i - d_i A = \partial_t F_i - \partial_i A + A F'_i - F_i A'.
\end{equation}

With these definitions, the Einstein equations take the form 
\begin{subequations}\label{eq:Einseqs}
\begin{align}
0 &= \text{tr}\!\left(G'' - \tfrac{1}{2} G'^2\right) 
+ (n+1) \left( \frac{W''}{W} - \tfrac{1}{2} \frac{W'^2}{W^2} \right), \\[6pt]
0 &= A'' + \tfrac{1}{2} \tilde{\nabla}\cdot F' + \tfrac{1}{2} F'\cdot F' 
+ \tfrac{1}{2} (\text{tr } d_+ G)' + \tfrac{1}{4} \text{tr}(G' d_+ G) 
+ \frac{2\Lambda}{D-2} \nonumber \\ 
&\quad + (n+1) \left( \tfrac{1}{4} \frac{\tilde{\nabla} W}{W}\cdot F' 
+ \tfrac{1}{2} \Big(\tfrac{d_+ W}{W}\Big)' 
+ \tfrac{1}{4} \frac{W' d_+ W}{W^2} \right),\\[6pt]
0 &= \text{tr}\!\Big[d_+(d_+ G) - A'(d_+ G) - \tfrac{1}{2} (d_+ G)^2\Big] 
+ 2 \tilde{\nabla}\cdot E + \tfrac{1}{2} \text{tr}(\Omega^2) \nonumber \\
&\quad + (n+1) \left( \frac{d_+ d_+ W}{W} - \frac{A' d_+ W}{W} 
- \tfrac{1}{2} \frac{(d_+ W)^2}{W^2} + \frac{\tilde{\nabla} W}{W}\cdot E \right), \\[6pt]
0 &= F''_i - F'^k G'_{ki} + \tfrac{1}{2} (\text{tr } G') F'_i - \tilde{\nabla}^k G'_{ki} 
+ \tilde{\nabla}_i(\text{tr } G') \nonumber \\ 
&\quad + (n+1) \left[ \tfrac{1}{2} \tfrac{W'}{W} F'_i 
- \tfrac{1}{2}\Big( \tfrac{\tilde{\nabla}_k W}{W} G'^k_i + \tfrac{W'}{W^2} \tilde{\nabla}_i W \Big) 
+ \tfrac{\tilde{\nabla}_i W'}{W} \right], \\[6pt]
0 &= d_+ F'_i + \tilde{\nabla}_k(d_+ G)^k_i - \tilde{\nabla}_i(\text{tr } d_+ G) 
+ \tfrac{1}{2} (\text{tr } d_+ G) F'_i - 2 \tilde{\nabla}_i A' - G'^k_i E_k \nonumber \\
&\quad + \tilde{\nabla}_k \Omega^k{}_i + F'_k \Omega^k{}_i \nonumber \\
&\quad + (n+1) \left[ \tfrac{1}{2}\left( \tfrac{\tilde{\nabla}_k W}{W} (d_+ G)^k{}_i 
+ \tfrac{d_+ W}{W^2} \tilde{\nabla}_i W \right) 
- \tfrac{\tilde{\nabla}_i d_+ W}{W} + \tfrac{1}{2} \tfrac{d_+ W}{W} F'_i 
+ \tfrac{1}{2} \tfrac{\tilde{\nabla}_k W}{W} \Omega^k{}_i \right], \\[6pt]
0 &= \Big\{ (d_+ G_{ij})' - G'^k_i d_+ G_{kj} 
+ \tfrac{1}{4} (\text{tr } G') d_+ G_{ij} 
+ \tfrac{1}{4} G'_{ij} \text{tr}(d_+ G) \nonumber \\
&\quad - \tilde{R}_{ij} + \tfrac{2\Lambda}{D-2} G_{ij} + \tilde{\nabla}_i F'_j 
+ \tfrac{1}{2} F'_i F'_j \nonumber \\
&\quad + (n+1) \left[ \tfrac{1}{4} \tfrac{W'}{W} d_+ G_{ij} 
+ \tfrac{1}{4} \tfrac{d_+ W}{W} G'_{ij} 
+ \tfrac{1}{2}\left( \tfrac{\tilde{\nabla}_i \tilde{\nabla}_j W}{W} 
- \tfrac{\tilde{\nabla}_i W \tilde{\nabla}_j W}{2 W^2} \right)\right]\Big\} + (i \leftrightarrow j), \\[6pt]
0 &= (d_+ W)' + \tfrac{1}{2} \tilde{\nabla}^2 W + \tfrac{1}{4} (\text{tr } G') d_+ W 
+ \tfrac{1}{4} (\text{tr } d_+ G) W' + \tfrac{1}{2} F'^k \tilde{\nabla}_k W 
+ \frac{2\Lambda}{D-2} W \nonumber \\
&\quad - \kappa n + \tfrac{n-1}{2} \left[ \tfrac{W' d_+ W}{W} 
+ \frac{(\tilde{\nabla} W)^2}{2W} \right]. 
\end{align}
\end{subequations}

\section{Large-$D$ limit of background geometries}\label{app:rescSigma}

In this appendix we present the most common background metrics in the large-$D$ regime, with radial distances of $O(1/D)$ and membrane lengths of $O(1/\sqrt{D})$, in the Bondi-Sachs form used in the text. They are all employed in Section~\ref{sec:examples}.

Consider the class of spacetimes
\begin{align}\label{eq:MAdS}
    ds^2=-(\kappa-\sigma r^2)dt^2+2dt\,dr+r^2 d\Sigma_{D-2}^{(\kappa)}\,,
\end{align}
where we recall that $\Sigma_{D-2}^{(\kappa)}$ is a $(D-2)$-dimensional space of constant curvature $\kappa=+1,0,-1$.
These spacetimes are Minkowski ($\sigma=0,\,\kappa=1$), AdS ($\sigma=-1,\,\kappa=\pm 1,0$) and dS ($\sigma=1,\,\kappa=1$) in Bondi-Sachs (or Eddington-Finkelstein) coordinates.

To take the appropriate large-$D$ limit, we write
\begin{equation}\label{eq:bckgs0}
    d\Sigma_{D-2}^{(\kappa)}=d\xi^2+ s_\kappa^2(\xi)\,d\Omega_{p-1} +c_\kappa^2(\xi)\, d\Sigma_{n+1}^{(\kappa)}\,,
\end{equation}
where for $\kappa=+1,0,-1$ we use
\begin{equation}
    s_\kappa(\xi) =\sin\xi\,,\xi\,,\sinh\xi\,,\qquad
    c_\kappa(\xi) =\cos\xi\,,1\,,\cosh\xi\,,
\end{equation}
which satisfy $c_\kappa^2(\xi)+\kappa s_\kappa^2(\xi)=1$.
As usual, we keep $p$ finite as $D,n\to\infty$.  Rescaling now $\xi=x/\sqrt{D}$ and retaining only terms of leading order in $1/D$ we obtain
\begin{equation}
    d\Sigma_{D-2}^{(\kappa)}\to \frac{\delta_{ij}}{D}dx^i dx^j+\left(1-\kappa\frac{x^i x_i}{D}\right) d\Sigma_{n+1}^{(\kappa)}\,,
\end{equation}
where $x^i$, $i=1,\dots,p$ are cartesian coordinates centered around $\xi=0$ up to a distance $\sim 1/\sqrt{D}$. The term $\propto 1/D$ in the coefficient multiplying $d\Sigma_{n+1}^{(\kappa)}$ must be retained since it exponentiates to a power $n+1$ in the field equations and thus becomes $O(D^0)$.

Next we introduce the radial coordinate $\rho$ in \eqref{eq:rrho}, rescale $t\to t/r_0$, and take $D\to\infty$. To leading order in $1/D$, \eqref{eq:MAdS} becomes
\begin{equation}\label{eq:bckgs}
    ds^2=
    \frac{\delta_{ij}}{D}dx^i dx^j-\lambda^2
    dt^2+\frac{2}{D}dtd\rho + \left(1+\frac{2\rho}{D}-\kappa\frac{x^i x_i}{D}\right)d\Sigma^{(\kappa)}_{n+1}\,,
\end{equation}
with $\lambda$ defined in \eqref{eq:deflambda}. In terms of the sources in \eqref{eq:sources}, this background has 
\begin{equation}\label{eq:gens}
    s(x)=-\kappa\frac{x^i x_i}2\,.
\end{equation}

In Minkowski, the source $s$ can be avoided if, instead of \eqref{eq:MAdS}, we write flat spacetime as 
\begin{align}
    ds^2=\delta_{ij}d\xi^i d\xi^j -dt^2+2dt\,dr + r^2 d\Omega_{n+1}\,.
\end{align}
Then, when $D\to\infty$,
\begin{align}\label{eq:Mink}    
     ds^2\to \frac{\delta_{ij}}{D}dx^i dx^j  -dt^2+\frac{2}{D}dtd\rho+
    \left(1+\frac{2\rho}{D}\right)d\Omega_{n+1}\,.
\end{align}
The form \eqref{eq:bckgs} is the background employed to describe the Schwarzschild black hole in Section~\ref{subsec:schlD}, while \eqref{eq:Mink} is used in Section~\ref{subsec:schblob}---the backgrounds appear as the asymptotic geometries in the limit $1\ll \rho\ll n$.

Finally, we discuss the backgrounds appropriate for the AdS black strings \cite{Emparan:2021ewh,Licht:2022rke}, namely AdS$_D$ with AdS$_{D-1}$ slices,
\begin{align}
    ds^2=\frac1{\cos^2 z}\left( dz^2-(r^2+\kappa)dt^2 +2 dt\,dr + r^2 d\Sigma^{(\kappa)}_{D-3}\right)\,,
\end{align}
with $z\in(-\pi/2,\pi/2)$. We focus on the small $z$ region, with $z=x/\sqrt{D}$, then change to the radial coordinate $\rho$ in \eqref{eq:rrho}, rescale $t\to t/r_0$, and finally take the large-$D$ limit to leading order as explained above. We obtain
\begin{align}\label{eq:hypads}    
     ds^2\to \frac{dx^2}{D}-\lambda^2 dt^2+\frac{2}{D}dtd\rho+ \left(1+\frac{2\rho}{D}+\frac{x^2}{D}\right)d\Sigma^{(\kappa)}_{D-3}\,,
\end{align}
with $\lambda^2=1+\kappa/r_0^2$.
Now the source is
\begin{equation}\label{eq:hypadss} 
    s(x)=\frac{x^2}2\,.
\end{equation}
These backgrounds are used in Section~\ref{subsec:adsbstring}.

\section{Technical details for Section~\ref{sec:HfromM}}\label{app:techdet}

\subsection{Geometrical quantities: Background}
Extrinsic curvature:
\begin{align}
    K_{tt}&=\sigma\lambda+O(1/D),\\
    K_{t\rho}&=K_{\rho t}=-\frac{\sigma}{\lambda\,D}+O(1/D^2),\\
    K_{\rho\rho}&=\frac{\sigma}{\lambda^3\,D^2}+O(1/D^3),\\
    K_{i\rho}&=K_{\rho i}=O(1/D^2),\\
    K_{tj}&=K_{jt}=\frac{1}{\lambda\,D}\left(-\nabla_j \dot{y}+\left(\nabla^i y-f^i\right)\left(\frac{1}{2}\dot{\gamma}_{ij}+\nabla_{[i}f_{j]}\right)+\nabla_j A^{(1)}-\frac{\partial_\rho F^{(1)}_j}{2}\right)+O(1/D^2),\\
    K_{ij}&=\frac{1}{\lambda\,D}\left(-\nabla_i\nabla_j y+\nabla_{(j}f_{i)}+\frac{1}{2}\left(\dot{\gamma}_{ij}+\partial_\rho G^{(1)}_{ij}\right)\right)+O(1/D^2)\,,
\end{align}
and along the directions of $\Sigma_{n+1}^{(\kappa)}$ :
\begin{align}
     K_{ab}&=r_0^2\gamma_{ab}\lambda\nn\\
     &+\frac{r_0^2\gamma_{ab}}{\lambda\,D}\left(\lambda^2y+2\lambda^2 s+s_a^{(1)}+A^{(1)}+\dot{s}+(f-\nabla y)\cdot \left(\nabla s+\frac{f+\nabla y}{2}\right)\right)+O(1/D)   
\end{align}
with $\gamma_{ab}$ the metric of $\Sigma_{n+1}^{(\kappa)}$.

Extrinsic curvature trace:
\begin{align}
    \lambda\,K=&\lambda^2\,D
    +s_a^{(1)}+A^{(1)}-\lambda^2(y+p+2)+\dot{s}+\left(\nabla+\nabla s+\frac{f+\nabla y}{2}\right)\cdot \left(f-\nabla y\right)\nn\\&+\frac{1}{2}\textrm{tr}\left(\dot{\gamma}+\partial_\rho G^{(1)}\right)+\sigma+O(1/D).
\end{align}

\subsection{Geometrical quantities: Membrane}
Velocity field:
\begin{align}
    u^t&=\frac{1}{\lambda},\\
    u^j&=\frac{v^j}{\lambda},\\
    u_t&=g_{tt}u^t+g_{ti}u^i=-\lambda+O(1/D),\\    u_j&=g_{jt}u^t+g_{ij}u^i=\frac{1}{\lambda\,D}\left(v_j+\nabla_j y-f_j\right)+O(1/D^2).
\end{align}
Extrinsic curvature:
\begin{align}       
K_{\rho\rho}&=O(1/D^2),\\
K_{\rho t}&=O(1/D),\\
K_{\rho j}&=O(1/D^2),\\
\mathcal{K}_{ab}&=K_{ab},\\
K_{tt}&=O(1)\rightarrow \mathcal{K}_{tt}=K_{tt}+O(1/D),\\
K_{ij}&=O(1/D)\rightarrow \mathcal{K}_{ij}=K_{ij}+O(1/D^2),\\
K_{jt}&=O(1/D)\rightarrow \mathcal{K}_{tj}=K_{tj}+\partial_j y K_{yt}+O(1/D^2),
\end{align}
where $a,b$ are indices along $\Sigma_{n+1}^{(\kappa)}$.

\subsection{Projector term}
\begin{align}
    \mathcal{P}^\mu _j \mathcal{V}_\mu &= \mathcal{V}_a+u_j u\cdot \mathcal{V}=\mathcal{V}_j+\frac{v_j+\nabla_j y-f_j}{\lambda\,D}\left(\mathcal{V}_t+v^i\mathcal{V}_i\right),\\
    \mathcal{P}^i_j&=\delta^i_j+u_ju^i=\delta^i_j+\frac{v^i}{\lambda\, D}\left(v_j+\nabla_j y-f_j\right)=\delta^i_j+O(1/D)=O(1),\\
    \mathcal{P}^t_j&=u^t u_j=\frac{v_j+\nabla_j y-f_j}{\lambda\, D}+O(1/D^2)=O(1/D),\\
    \mathcal{P}^t_j u^\mu \mathcal{K}_{\mu t}&=\frac{\sigma}{\lambda^2\,D}\left(v_j+\nabla_j y-f_j\right)+O(1/D^2).
\end{align}

\subsection{Vector equation simplification}\label{app:veceqn}

Using the expressions in \eqref{eq:vecterms}, the vector equation \eqref{eq:vectorM} becomes
\begin{align}
    0=&\left(\nabla^i+\nabla^i s+\nabla^i y-v^i\right)\nabla_i v_j -\dot{v}_j
    +v^i\left(-\nabla_i\nabla_j s+R_{ij}+2\nabla_{[i}f_{j]}\right)\nn\\
    &~ +\left(\nabla^i s+\nabla^i y-v^i\right)\nabla_i\nabla_j y+\nabla_j\nabla^2 y+(\lambda^2+2\sigma)\nabla_j y-\nabla_j \dot{y}\nn\\
    &~ +\nabla^i y\left(\nabla_i\nabla_j s+\dot{\gamma}_{ij}\right)
    +\text{Sources}\,,
\end{align}
where the sources, which include all terms independent of both $v_j$ and $y$, are
\begin{align}  
\text{Sources}&=\nabla_j\left(2(\lambda^2+\sigma)s+s_a^{(1)}+\nabla^2 s+(\nabla s)^2+f\cdot \nabla s+\frac{1}{2}f^2-\frac{1}{2}\text{tr}(\dot{\gamma})\right)\nn\\
    &\quad +\dot{f}_j+\left(\nabla^i+\nabla^i s\right)\dot{\gamma}_{ij}\nn\\
    &=  \nabla_j f_0+\dot{f}_j+\left(\nabla^i+\nabla^i s\right)\dot{\gamma}_{ij}.
\end{align}
Now we change $v_j= \hat{p}_j/m-\nabla_jy$ and conveniently rearrange it as
\begin{align}
    0=&-\frac{\hat{p}^i}{m}\nabla_i \frac{\hat{p}_j}{m}-\partial_t{\left(\frac{\hat{p}_j}{m}\right)}+\frac{\hat{p}^i}{m}\left(-\nabla_i\nabla_j s+R_{ij}+2\nabla_{[i}f_{j]}\right)\nn\\
    &+\left(\nabla^i+\nabla^i s+2\nabla^i y\right)\nabla_i\frac{\hat{p}_j}{m}    +\left(\lambda^2+2\sigma\right)\nabla_j y\nn\\
    &+\nabla^i y\left(2\nabla_i\nabla_j s-2R_{ij}-2\nabla_{[i}f_{j]}+\dot{\gamma}_{ij}\right)
    +\left(\nabla^i+\nabla^i s\right)\dot{\gamma}_{ij}+\dot{f}_j+\nabla_j f_0.
\end{align}
Further changing to $y=\log m$, we get
\begin{align}
    0=&-\frac{\hat{p}^i}{m}\nabla_i \hat{p}_j+\frac{\hat{p}_j\hat{p}_i}{m^2}\nabla^i m-\partial_t{\hat{p}_j}+\frac{\partial_t{m}}{m}\hat{p}_j+\hat{p}^i\left(-\nabla_i\nabla_j s+R_{ij}+2\nabla_{[i}f_{j]}\right)\nn\\
    &+\left(\nabla^i+\nabla^i s\right)\nabla_i\hat{p}_j-\hat{p}_j\left(\frac{\nabla^2 m}{m}+\nabla^i s\frac{\nabla_i m}{m}\right)    +\left(\lambda^2+2\sigma\right)\nabla_j m\nn\\
    &+\nabla^i m\left(2\nabla_i\nabla_j s-2R_{ij}-2\nabla_{[i}f_{j]}+\dot{\gamma}_{ij}\right)
    +m\left(\nabla^i+\nabla^i s\right)\dot{\gamma}_{ij}+m\dot{f}_j+m\nabla_j f_0.
\end{align}    
We can eliminate $\partial_t m$ using the mass equation \eqref{eq:effm}, to find
\begin{align}
    0=&-\frac{\hat{p}^i}{m}\nabla_i \hat{p}_j+\frac{p_jp^i}{m}\nabla_i m-\frac{\hat{p}_j}{m}\nabla^i \hat{p}_i-\nabla^i s\frac{\hat{p}_i \hat{p}_j}{m}-\partial_t{\hat{p}}_j-\left(\frac{1}{2}\text{tr}(\dot{\gamma})+\dot{s}\right)\hat{p}_j \nn\\
    &+\hat{p}^i\left(-\nabla_i\nabla_j s+R_{ij}+2\nabla_{[i}f_{j]}\right)
    +\left(\nabla^i+\nabla^i s\right)\nabla_i\hat{p}_j
    +\left(\lambda^2+2\sigma\right)\nabla_j m\nn\\
    & +\nabla^i m\left(2\nabla_i\nabla_j s-2R_{ij}-2\nabla_{[i}f_{j]}+\dot{\gamma}_{ij}\right)+m\left(\nabla^i+\nabla^i s\right)\dot{\gamma}_{ij}+m\dot{f}_j+m\nabla_j f_0\,,
\end{align}
which can be regrouped into the form of \eqref{eq:effp}.

\bibliography{refs}

\end{document}